# Improved Robustness for Deep Learning-based Segmentation of Multi-Center Myocardial Perfusion MRI Datasets Using Data Adaptive Uncertainty-guided Space-time Analysis

Dilek M. Yalcinkaya, MS [1,2], Khalid Youssef, PhD [1,3], Bobak Heydari, MD [4], Janet Wei, MD [5], Noel Bairey Merz, MD [5], Robert Judd, PhD [6], Rohan Dharmakumar, PhD [3,7], Orlando P. Simonetti, PhD [8], Jonathan W. Weinsaft, MD [9], Subha V. Raman, MD [3,10], Behzad Sharif, PhD [1,2,3,7,*]

[1] Laboratory for Translational Imaging of Microcirculation, Indiana University School of Medicine, Indianapolis, IN, USA

[2] Elmore Family School of Electrical and Computer Engineering, Purdue University, West Lafayette, IN, USA

[3] Krannert Cardiovascular Research Center, Dept. of Medicine, Indiana Univ. School of Medicine, Indianapolis, IN, USA

[4] Stephenson Cardiac Imaging Centre, Department of Cardiac Sciences, University of Calgary, Alberta, Canada

[5] Barbra Streisand Women's Heart Center, Smidt Heart Institute, Cedars-Sinai Medical Center, Los Angeles, CA, USA

[6] Division of Cardiology, Department of Medicine, Duke University, Durham, NC, USA

[7] Weldon School of Biomedical Engineering, Purdue University, West Lafayette, IN, USA

[8] Department of Medicine, Davis Heart and Lung Research Institute, The Ohio State University, Columbus, OH, USA

[9] Division of Cardiology at NY Presbyterian Hospital, Weill Cornell Medical Center, New York, NY, USA

[10] OhioHealth, Columbus, OH, USA

* Correspondence to:

Behzad Sharif, PhD

Laboratory for Translational Imaging of Microcirculation

Indiana University School of Medicine

1800 N Capitol Ave, Suite E371, Indianapolis, IN, USA 46202

Phone: (317) 274-0943 | Email: bsharif@iu.edu

## ABSTRACT

**Background.** Fully automatic analysis of myocardial perfusion MRI datasets enables rapid and objective reporting of stress/rest studies in patients with suspected ischemic heart disease. Developing deep learning techniques that can analyze multi-center datasets despite limited training data and variations in software (pulse sequence) and hardware (scanner vendor) is an ongoing challenge.

**Methods.** Datasets from 3 medical centers acquired at 3T (n = 150 subjects; 21,150 first-pass images) were included: an internal dataset (inD; n = 95) and two external datasets (exDs; n = 55) used for evaluating the robustness of the trained deep neural network (DNN) models against differences in pulse sequence (exD-1) and scanner vendor (exD-2). A subset of inD (n = 85) was used for training/validation of a pool of DNNs for segmentation, all using the same spatiotemporal U-Net architecture and hyperparameters but with different parameter initializations. We employed a space-time sliding-patch analysis approach that automatically yields a pixel-wise "uncertainty map" as a byproduct of the segmentation process. In our approach, dubbed Data Adaptive Uncertainty-Guided Space-time (DAUGS) analysis, a given test case is segmented by all members of the DNN pool and the resulting uncertainty maps are leveraged to automatically select the "best" one among the pool of solutions. For comparison, we also trained a DNN using the established approach with the same settings (hyperparameters, data augmentation, etc.).

**Results.** The proposed DAUGS analysis approach performed similarly to the established approach on the internal dataset (Dice score for the testing subset of inD: $0.896 \pm 0.050$ vs. $0.890 \pm 0.049$; p = n.s.) whereas it significantly outperformed on the external datasets (Dice for exD-1: $0.885 \pm 0.040$ vs. $0.849 \pm 0.065$, $p < 0.005$; Dice for exD-2: $0.811 \pm 0.070$ vs. $0.728 \pm 0.149$, $p < 0.005$). Moreover, the number of image series with "failed" segmentation (defined as having myocardial contours that include bloodpool or are noncontiguous in $\geq 1$ segment) was significantly lower for the proposed vs. the established approach (4.3% vs. 17.1%, $p < 0.0005$).

**Conclusions.** The proposed DAUGS analysis approach has the potential to improve the robustness of deep learning methods for segmentation of multi-center stress perfusion datasets with variations in the choice of pulse sequence, site location or scanner vendor.

## INTRODUCTION

In the past decade, landmark multi-center studies have established the clinical impact and cost effectiveness of stress/rest perfusion cardiac magnetic resonance (CMR) imaging with visual (expert) assessment in diagnosis and risk assessment of patients with ischemic heart disease (1-3). In the setting of a multi-center study, it would be desirable to enable objective, automated assessment of myocardial perfusion imaging data to avoid reader bias and/or to enable innovative retrospective analysis of multi-center data (4). This need is more evident in the setting of clinical trials or large international registry studies involving cardiovascular imaging (5) where it is costly, impractical, or undesirable to employ a uniform scanner/hardware platform and identical data acquisition protocols.

Recently, the potential for ubiquitous adoption of artificial intelligence (A.I.) algorithms in CMR has emerged (6-11) thanks to technical advances in the field of deep learning that may enable comparable performance to expert readers while eliminating the manual, time-consuming steps needed for reporting CMR examinations (12). However, ensuring effective generalization is a critical component of deep learning-based methods before they can be deployed in clinical practice (13). In this context, deep neural nets (DNNs) trained with limited data need to have a sufficient level of robustness to "dataset shifts," i.e., changes in the characteristics of the external dataset, as it is infeasible to acquire and incorporate data from every possible combination of MR scanner platform (hardware and software variations) during model training/development. Here, data augmentation (14, 15) or "data adaptive" approaches (16) can mitigate the impact of dataset shifts which are common in multi-center CMR studies.

Certain classes of DNN-based methods can provide a form of "generalization check" for trained models applied to external datasets (13), e.g., a measure of confidence for the generated image segmentation contours. Methods that generate a measure of uncertainty for the DNN output (17-20) fall into this category and have been studied in the context of CMR segmentation (21-25, 53). Of note, previous works have shown the effectiveness of quality control measures in DNN-based segmentation of native T1 maps (21-23). Unlike native T1 mapping, however, dynamic contrast-enhanced perfusion CMR datasets can have temporal characteristics that vary widely depending on the protocol (pulse sequence, contrast dose/agent, choice of stress agent), scanner hardware, patient cohort, etc. To enable a measure of "generalization check" for DNN-based analysis of external perfusion CMR datasets, we have recently proposed a space-time approach for uncertainty estimation that uses patch-level analysis to generate a pixel-wise map of segmentation uncertainty for a given DNN model by evaluating the agreement and/or inconsistency among outputs of the same (fixed) model among overlapping patches (24, 25).

This approach automatically creates an uncertainty assessment tool for any previously trained patch-level DNN, hence providing a form of "quality control" for DNN-based perfusion CMR analysis. Nevertheless, on its own, this setup would require the clinician reader to be involved in the process to provide expert input for uncertain (i.e., possibly incorrect) myocardial segments. In this proof-of-concept work, we propose a framework that incorporates a measure of space-time uncertainty during the analysis stage while eliminating additional effort by the reader. To this end, we introduce an uncertainty-guided technique for segmentation of perfusion CMR datasets with the goal of improving robustness of DNNs trained on limited single-center data to dataset shifts in a multi-center setting due to variations in scanner vendor or data acquisition protocol.

**METHODS**

**Internal and External Perfusion CMR Datasets**

The perfusion CMR datasets used in this retrospective study were obtained from vasodilator stress first-pass perfusion MRI scans in subjects with suspected myocardial ischemia, and were acquired at three sites: Cedars-Sinai Medical Center, Los Angeles, referred to as the internal Dataset (inD), University of Calgary, Canada, referred to as the 1st external dataset (exD-1), and Weill Cornell Medicine, New York City, referred to as the 2nd external dataset (exD-2). Local Institutional Review Board approval and written informed consent were obtained for all the volunteer and patient imaging studies before each imaging exam/study.

Table 1 summarizes the characteristics of the inD and the two external datasets (exD-1 and exD-2) used in this work. The inD (n = 95 stress/rest studies) was partitioned into three subsets: training dataset (n = 75), validation dataset (n = 10), and the internal "test dataset" (n = 10) referred to as inD-test. For evaluation of DNN-based automatic segmentation, three test datasets were used: inD-test, exD-1 (n = 40 stress-only studies), and exD-2 (n = 15 stress/rest studies). As highlighted in Table 1, there were significant demographic differences between inD and the external datasets, e.g., inD consisted of 90% female subjects whereas most of the study population in exD-1 and exD-2 were male. For all perfusion CMR studies, 3 short-axis slices were acquired and, as described in Table 1, 3T clinical scanners from two vendors were used: Siemens Healthineers (inD and exD-1) and GE Healthcare (exD-2). The rest/stress studies in inD were acquired using a vendor-provided saturation recovery (SR)-prepared pulse sequence with RF-spoiled gradient-recalled echo (GRE) readout (SR-prepared FLASH) on a Siemens 3T scanner (Magneton Verio). The studies in exD-1 were obtained on a different 3T Siemens scanner line (Magnetom Skyra) and an SR-prepared balanced steady-state free precession (bSSFP) pulse sequence. Data from exD-2 was acquired on a 3T GE scanner (Discovery MR750w) using a SR-prepared fast GRE sequence with different pulse sequence settings compared to inD as described in the caption of Table 1. All datasets used in this study were respiratory motion corrected using the vendor-provided (inline) motion correction feature.

**Overview of the Proposed Uncertainty-guided Space-time Analysis Approach**

Figure 1 (central illustration) describes an overview of the proposed approach, dubbed Data Adaptive Uncertainty-guided Space-time (DAUGS) analysis. In the training stage (top section of Fig. 1), we train a pool of fifty DNN models that have the same network architecture and same hyperparameters (learning rate, patch and stride size, etc.) but differ in the initialization of network weights/parameters. Inspired by prior deep-learning work in computational microscopy (26) and medical imaging (24, 25, 27, 28), we use a patch-level training approach and select the hyperparameters (fixed for the pool of DNNs) based on the validation dataset. In the analysis stage (bottom section of Fig. 1), a "test" image series (e.g., stress perfusion CMR time series from an external dataset) is analyzed by all models in the DNN pool. Each of the fifty analyses yields a segmentation solution and, as a byproduct (described in detail below), generates a corresponding pixel-wise "uncertainty map" (U-map). In the final step, the U-maps are leveraged to automatically choose the "best" result (the one with the lowest mean per-pixel uncertainty) among the pool of segmentation solutions (Step 4 in Fig 1). Further details for the training and analysis stages are provided in the following sections.

**Training Stage: Data Labeling, Preprocessing, and Data Augmentation**

The stress/rest CMR perfusion datasets from all three centers (Table 1) were meticulously segmented by two expert readers and the segmentation contours were imported using ITK-SNAP (29). The objective for the proposed DNN-based automatic segmentation framework is to classify each pixel in a given dynamic (2D+time) perfusion image series into the following three classes: left ventricular (LV) myocardium, bloodpool, or background. In all training/validation and testing datasets, the raw (scanner generated) first-pass perfusion image series were preprocessed by: (a) 2-fold spatial upsampling and automatic cropping (30) to the region of interest (ROI) centered around the heart location (128x128 spatial matrix); and (b) piecewise cubic interpolation for temporal resampling (30 time frames), resulting in a 128x128x30 spatiotemporal matrix. Subsequently, each perfusion image series was normalized to a pixel intensity range of [0, 1]. Figure 2 describes the various augmentation schemes that were employed during the DNN training process. Specifically, as described previously (24), two types of data augmentation were used: (a) segmentation-variant augmentation (spatial transforms that also change the corresponding ground-truth segmentation contours), and (b) segmentation-invariant augmentation that does not affect the ground-truth contours. For the latter, in addition to noise and contrast enhancement/degradation, the image series in the training dataset were augmented using spatial modulation with elaborately-designed intensity maps to induce various forms of coil-sensitivity weighting patterns (24). Further details for the data augmentation approach are provided in Appendix A (Supplementary Material).

**Training Stage: Patch-Level Training of the DNN Pool**

The concept of decomposing images into patches has been heavily used for segmenting images in computer vision applications (31, 32) and, more recently, in deep learning-based medical imaging applications (33-36) including perfusion CMR (28). As described in Fig 3(a), the input dynamic image series in our proposed framework is stacked into a 2D+time format and then decomposed into space-time patches by applying a spatially sliding window. Next, each space-time patch is independently analyzed by a trained DNN, which works "at patch-level," i.e., analyzes the space-time patch to detect the myocardial pixels. The segmented space-time patches (outputs of the DNN) are then combined to yield the segmentation solution. As model architecture, we used a "vanilla" 2D+time U-Net architecture (37) and optimized the cross-entropy loss using the Adam optimizer. Further details of the training process are provided in Appendix A.

A spatial sliding-window extracts the patches with a stride of half patch size (spatial) during training. Extensive numerical experiments were conducted to determine desirable choices for patch size and stride with details provided in Appendix B (Supplementary Material). At the output of the DNN, each space-time patch input resulted in static probability maps (also known as the softmax probability) of size 64×64 corresponding to the probability of each pixel belonging to one of the three classes of pixels (myocardium, bloodpool, background). Fig. 3(a) summarizes the data processing pipeline for the proposed patch-level DNN. Thanks to the patch-level sliding window approach, a specific pixel in the 2D+time perfusion images is processed multiple times since it belongs to multiple overlapping patches. To this end, myocardial softmax probabilities (segmentation probability maps) were averaged for pixels that are in multiple patches. In this way, during inference, DNN-segmented patches are combined back with an overlap of one-fourth patch size to yield the segmentation solution. Spatial sliding window extracts more patches from the center of the image series than the edges, which is convenient given that the images are localized around the heart with the ROI in the center. Additional details for patch-level training and analysis are provided in Appendix B. The deep learning work in this study were implemented using MATLAB (R2020b; The MathWorks Inc., Natick, MA) and trained on an NVIDIA Titan RTX graphical processing unit with 24 GB onboard memory.

**Analysis Stage: Sliding-patch Uncertainty Estimation**

Figure 3(b) provides a pictorial description of how patch-level analysis enables automatic computation of a pixel-wise uncertainty map (U-map). With a small step size (stride) for the patch sliding-window, each pixel in the perfusion image series belongs to multiple patches and therefore is segmented multiple times during the analysis stage. Ideally a pixel that belongs to, e.g., five overlapping space-time patches (orange-shaded volume in Fig 3b) should be classified in the same way by the DNN in all five patches; therefore, the level of discrepancy between the output of the DNN (when applied to overlapping patches) is indicative of reduced confidence (higher uncertainty) in DNN-based analysis. Here we provide a condensed mathematical description of how the U-map and our uncertainty metric are computed automatically as a byproduct of the patch-level segmentation process. Briefly, for a given perfusion image series in the analysis stage as visualized in Fig 3(b), let $\Gamma_{(x,y)}$ denote the set of space-time patches that include a particular pixel location $(x,y)$ in the 2D+time image series. For the $i$-th patch in $\Gamma_{(x,y)}$, let $p^i(x,y)$ denote the probability scores computed by a trained DNN during inference, indicating the likelihood of pixel location $(x,y)$ belonging to the LV myocardium. The U-map is obtained by computing the discrepancy between patch-level probability scores $p^i$ across the entire ROI (25):

$$\text{U-map}(x,y) = \texttt{std}(p^1_{(x,y)}, p^2_{(x,y)}, \dots, p^{|\Gamma_{(x,y)}|}_{(x,y)})$$

where `std` is the standard deviation operator. Here, we define the following metric to quantify the mean per-pixel energy in the U-map: $U_{\text{pp}} = ||U-\text{map}||_F^2 / N_{myo}$ where $||\cdot||_F$ denotes the Frobenius norm and $N_{myo}$ is the number of myocardial pixels in the segmentation solution (Appendix C in Supplementary Materials provides further details). In other words, $U_{pp}$ for a particular segmentation solution is the mean value of the sum-of-squares of its corresponding U-map "image". It is shown in Appendix C that the range of the U-map values is in the [0, 0.5] interval (which is reflected in how the maps are displayed in Results). Our previous work demonstrated the usefulness and validity of the U-map concept in improving the interpretation of DNN-derived patch-level segmentation of CMR perfusion datasets (24, 25). To complement this prior work, in Appendix D (Supplementary Material), we have described a simulation experiment showing that the uncertainty metric used in our proposed approach ($U_{pp}$) can track the level of uncertainty (difficulty) for myocardial segmentation in the presence of varying levels of nonrigid motion-correction errors.

**Analysis Stage: Data Adaptive Uncertainty-Guided Space-time (DAUGS) Analysis**

As outlined above (Fig. 1), the goal of the proposed DAUGS analysis methodology is to improve robustness in segmenting myocardial (endo/epi) contours of multi-center multi-vendor perfusion CMR datasets. During DAUGS analysis of a particular case from a test dataset, each DNN in the model pool generates a segmentation solution and a corresponding U-map (Step 3 in Fig. 1). Next, the segmentation solution that has the lowest mean per-pixel energy in its U-map ($U_{pp}$) is selected as the final segmentation solution (Step 4 in Fig. 1). This approach enables data-adaptive "model selection" on the basis of the uncertainty (quantified by $U_{pp}$) that a given test case induces in each model. The pool of trained DNNs for the proposed approach is created by running the training process for the 2D+time U-Net models five times, each time with a different parameter initialization. This is inspired by prior work showing that multiple random initializations of the DNN parameters (model weights) help with exploring different local minima in the "loss landscape" during the stochastic optimization process for training a DNN (41, 45). For each of the 5 training runs, we selected 10 checkpoints (snapshots of the working DNN model during the training process) that have ≥ 0.87 Dice score (7, 8) on the internal validation dataset for inclusion in the pool of trained DNNs (46). This resulted in a total of fifty DNNs in the trained model pool for the proposed DAUGS approach.

**Performance Evaluation: DAUGS Analysis vs. The Established Approach**

We compared the proposed DAUGS analysis versus the "established approach" for DNN-based image analysis. Specifically, in the established approach, among the same pool of trained DNN models (top section of Fig. 1), one DNN model was selected based on its performance on the validation dataset, i.e., the DNN with the best performance (highest

Dice score) on the validation dataset was selected and used to analyze the test datasets. In contrast, for our proposed data-adaptive approach (DAUGS analysis), once the DNN pool is trained using the training/validation data, in the analysis stage we let the test data inform which DNN model (among the pool of trained models) should be selected based on the associated uncertainty metric (bottom section of Fig 1).

The accuracy of automatic DNN-based segmentations on the test datasets (inD-test, exD-1, and exD-2) was evaluated using pixelwise Dice coefficient and undirected 95th-percentile Hausdorff distance (38). Differences in performance between the proposed vs. the established approaches were compared using unpaired Student t-tests. In addition to this form of quantitative performance comparison, we performed visual inspection of the segmentation results across the two external datasets (exD-1 and exD-2) to identify the number of "failed" segmentations according to the following criteria: (i) inclusion of bloodpool inside the delineated myocardial contours; (ii) noncontiguous myocardial contours in at least 1 myocardial segment. The difference between the two approaches in terms of the prevalence of failed segmentations were compared using Fisher's exact test. To evaluate the impact of segmentation accuracy on fully quantitative analysis of stress perfusion CMR, myocardial blood flow (MBF) quantification was performed on the two test datasets that included dual-bolus acquisition (inD-test and exD-1). Agreement and correlation between MBF derived based on the two automatic segmentation methods (DAUGS analysis and the established approach) vs. MBF derived from manual segmentation was evaluated using linear regression and Bland–Altman analysis. Further details on MBF quantification methodology are described in Appendix E (Supplementary Material). All statistical tests were two-tailed with a p-value < 0.05 considered statistically significant. Bivariate correlations were assessed using the Pearson coefficient. Continuous variables are described as mean ± standard deviation.

**RESULTS**

**Segmentation Performance: Internal Dataset vs. External Datasets**

Figure 4 summarizes the results comparing the overall myocardial segmentation performance of the proposed DAUGS analysis approach vs. the established DNN-based analysis approach across the three test datasets (one internal and two external). On the internal dataset (inD-test), the proposed approach resulted in a slightly higher myocardial segmentation accuracy compared to the established approach, however this overperformance was not significant (Dice

score: 0.896 ± 0.050 vs. 0.890 ± 0.049 and Hausdorff distance: 1.854 ± 0.267 mm vs. 1.867 ± 0.223 mm; p = n.s. for both comparisons). In contrast, for the first external dataset (exD-1; acquired with a different pulse sequence), DAUGS analysis significantly outperformed the established approach (Dice score: 0.885 ± 0.040 vs. 0.849 ± 0.065 and Hausdorff distance: 2.074 ± 0.297 mm vs. 2.211 ± 0.322 mm; p < 0.005 for both comparisons). Likewise, on the second external dataset (exD-2; acquired using a scanner from another vendor), DAUGS analysis resulted in significantly more accurate segmentation vs. the established approach (Dice score: 0.811 ± 0.070 vs. 0.728 ± 0.149 and Hausdorff distance: 1.931 ± 0.263 mm vs. 2.271 ± 0.370 mm; p < 0.005 for both comparisons).

According to the abovementioned definition for failed segmentation based on visual inspection, neither method had any failed segmentations in the internal dataset (inD-test); however, the number of perfusion image series with failed segmentation was significantly lower for the proposed vs. the established approach (4.3% vs. 17.1% of the total number of perfusion image series; p = 0.0003 for this comparison).

**Representative Cases from the External Datasets**

Figures 5-8 present the segmentation results and the accompanying uncertainty maps for three representative cases from exD-1 (Figs. 5, 6, and 7 corresponding to normal stress exam, focal stress-induced defect, and diffuse stress-induced defects, respectively) and one case from exD-2 (Fig. 8). In each figure, the stress perfusion CMR image series are shown at three stages of the first pass of the contrast agent (right/left bloodpool enhancement followed by myocardial enhancement) for three short-axis slices (basal, mid, and apical) together with manual (ground truth) contours and DNN-based automatic segmentation results. Note that the results for the established DNN-based analysis approach also have an accompanying uncertainty map. In Fig. 5, which corresponds to a normal stress exam, the DAUGS analysis-derived myocardial contours accurately segment all three slices whereas the results using the established approach have lower Dice scores and also show a failed (noncontiguous) segmentation for the apical slice (magenta arrow). Importantly, this error is reflected in the corresponding U-map and its mean per-pixel energy ($U_{pp}$). Figure 6 shows the results for a patient with focal stress-induced perfusion defect (single-vessel disease). Here, both approaches perform well.

Figure 7 shows a patient with diffuse stress-induced ischemia in all three short-axis slices and LV hypertrophy. The mid slice (acquired at end systole) is challenging to segment due to the difficulty in delineating the subendocardial/subepicardial borders especially in the septal region. The established DNN-based analysis approach performs notably worse in the apical slice and fails to segment the mid slice (indicated by yellow arrows) which is also reflected in the accompanying U-maps and $U_{pp}$ values. For this challenging case, the proposed DAUGS analysis method, however, performs well (mean Dice score of > 0.90 across three slices) while still demonstrating a high-level of uncertainty (see the corresponding U-map). Finally, Fig. 8 shows a representative case from exD-2 wherein the proposed approach demonstrates superior performance compared to the established approach which has a lower Dice score for basal and mid slices, in addition to a noncontiguous contour in the apical slice (highlighted by the yellow arrow). The magenta arrow (apical slice) points to the incorrect inclusion of epicardial fat layer in the segmentation solution by the established approach; this was successfully excluded by the proposed DAUGS analysis approach.

**Additional Evaluations and Observations**

Results of correlation and agreement analysis between MBF derived from segmentation solutions of the two automatic DNN-based methods and MBF derived from manual segmentation for inD-test and exD-1 are described in Appendix E and Suppl. Figures S2 and S3 (Supplementary Material). Briefly, the two methods performed similarly on the internal test set (inD-test) as shown in Suppl. Fig. S2; however, for the external dataset (exD-1), the proposed DAUGS analysis approach showed stronger correlation and tighter Bland-Altman limits of agreement vs. manual segmentation compared to the established approach as shown in Suppl. Fig. S3.

Next, we examined the heterogeneity of segmentation solutions in the DNN model pool (50 models) by visualizing all 50 segmentation solutions and their corresponding U-maps for two examples image series: one from exD-1 and one from exD-2. As described in Suppl. Fig. S4 and Appendix F, the observed matrix of 50 segmentation solutions shows noticeable heterogeneity/diversity in terms of quality of segmentation and U-map composition as well as the corresponding $U_{pp}$ metric when tested on the two external datasets. Finally, we evaluated an alternative to $U_{pp}$ as the uncertainty metric (specifically, the total energy of the U-map instead of the normalized per-pixel energy) by repeating the experiments when this alternative choice is used. The details are described in Appendix G (with an example result shown Suppl. Figure S5) and suggest that the normalization factor in $U_{pp}$ may be helpful in reducing the number of noncontiguous segmentations.

To further evaluate the generalization capability of the proposed method, we compared its performance on the external datasets (exD-1 and exD-2) to a DNN that did not suffer from "dataset shifts" (13), i.e., was trained on exD-1 and exD-2. Details are provided in Appendix H. Lastly, among the test cases, we encountered noticeable electrocardiogram mis-triggering in one of exD-1 cases wherein the MoCo algorithm failed to eliminate the nonrigid motion (due to abrupt changes in cardiac phase) for a handful of time frames. For this case, DAUGS analysis outperformed the established approach (mean Dice score across the 3 slices: $0.892 \pm 0.052$ vs. $0.853 \pm 0.048$; mean absolute MBF error for the mid slice with perfusion defect: $0.03 \pm 0.03$ vs. $0.06 \pm 0.03$).

**DISCUSSION**

In recent years, A.I.-enabled techniques have been applied for automatic analysis of stress/rest perfusion CMR datasets (7, 8, 28). Yet, there is currently no technique that can improve the generalizability of deep learning-based segmentation of perfusion CMR datasets for analyzing external datasets, e.g., in the setting of a multi-center study or clinical trial, where robustness to dataset shifts is needed despite variations in the data-acquisition platform, i.e., variations involving software (pulse sequence and protocol parameters) or hardware (scanner vendor). In this context, reliable segmentation of LV myocardium can be a key component of A.I.-assisted frameworks for streamlined reporting of stress perfusion studies (12) and/or objective quantification of ischemic burden based on "visual" assessment. Recent works in deep learning-based automatic MRI segmentation models suggest that the factors involved in successful multi-center, inter-institutional generalization of DNN models trained on datasets from a single center is not well understood even for musculoskeletal MRI for which publicly available multi-center training datasets exist (39). One approach to mitigate the performance loss typically observed when a trained DNN model is applied to external datasets is to "fine tune" the trained model to every new dataset, that is, to re-train an already trained DNN on a subset of the data from the external datasets (40). This fine-tuning process is costly, possibly prohibitively so, both in terms of additional effort needed to generate new training data (from external sites) as well as computational resources needed to perform the tasks involved.

In this work, we proposed a two-pronged strategy in the setting of multi-center perfusion CMR datasets to mitigate this performance loss without the need for fine-tuning as outlined in Fig. 1: (i) training a pool of models with identical setup (same architecture, training procedure, etc.) and different initial conditions that are expected to perform *similarly on average* (across a group of test cases) but may perform *differently on each particular test case*, and (ii) deriving a new uncertainty-guided quality metric that, during the analysis stage, enables data-adaptive selection of the best performing model from the pool of DNNs for each test case. Our results show that, compared to the established DNN-based analysis approach (i.e., picking the DNN model that performed the best among other models on the validation dataset), the proposed DAUGS analysis technique improves the segmentation performance for external datasets, i.e., in the presence of "dataset shifts" (13) caused by variations in MR scanner hardware (site location or scanner vendor) or the data acquisition settings (pulse sequence choice or protocol parameters). However, our approach did not significantly improve the segmentation performance on the internal test set (i.e., on the portion of inD not included in the training dataset), which is expected since, in the absence of dataset shifts the performance difference across the DNN pool should be minimal.

In this study, dataset shift or "data distribution" shift, a common phenomenon when deploying A.I. methods in multi-center settings (13), is encountered when moving from the internal dataset to the two external datasets. Besides the patient characteristic differences, (a) exD-1 uses a bSSFP readout whereas inD is acquired with a spoiled gradient-echo readout; (b) exD-2 is acquired on a different scanner platform (different vendor). As expected, the segmentation performance for the established approach deteriorated when moving from inD to the external datasets (mean Dice score for inD: 0.89 vs. 0.85 for exD-1 and 0.73 for exD-2). The proposed DAUGS approach also had a performance drop between the internal and external datasets; however, this drop was markedly lower for our technique (mean Dice score for inD: 0.90 vs. 0.89 for exD-1 and 0.81 for exD-2). Moreover, the number of image series with failed segmentation (noncontiguous contours or including bloodpool) was 4-fold lower for the proposed approach (4.3% vs. 17.1%, $p < 0.0005$). Furthermore, we assessed the impact of more accurate segmentation (higher Dice score and reduced failure rate) on the accuracy of automatic MBF quantification vs. manual analysis. We observed that, although the two methods perform similarly on the internal test set, the proposed approach has a higher level of agreement with respect to manual expert analysis on the external dataset (exD-1) when compared to the established approach (Appendix E). These results suggest that the proposed uncertainty-guided approach (DAUGS analysis technique) improves the robustness of DNN-based segmentation for external datasets by: (i) offering the flexibility to choose the final segmentation solution from a pool of candidate solutions (instead of a single solution) based on the detected uncertainty by DNNs; (ii) using the corresponding uncertainty score to select the best solution.

The latest technical developments in the field of deep learning, outside of medical imaging, have shown the potential advantage of employing a pool of DNNs or model ensembles, typically referred to as deep ensembles (41), trained using the same network architecture (hyperparameters) and the same training dataset but with variations in the initialization (random seed) for model weights. The key idea is that averaging the output of the DNN pool will "stabilize" the DNN-based analysis resulting in improved robustness and performance when tested on external datasets (42-44). More recently, there is an increased interest in developing "test time" data-adaptive techniques using a form of "generalization check" to improve the robustness of A.I. algorithms to dataset shifts (16). In our proposed method, DAUGS analysis, we combine these recent innovations by training the same DNN model five times independently, each with a different parameter initialization, and including the intermediate models obtained during the training process (10 for each of the 5 independent training runs) to create a pool of fifty DNN models (45, 46). Prior theoretical work in this area has shown that a DNN pool of this size can achieve sufficiently diverse predictions (in our case, segmentations) to choose from (41, 45). Next, in the analysis stage, our method computes a space-time uncertainty map (for the segmentation result of each member of the DNN pool) that is specifically suited for 2D+time (dynamic) CMR datasets, and coverts it to a generalization measure ($U_{pp}$) to adaptively choose a single model (among the 50-member DNN pool) for the particular perfusion CMR test-case being analyzed to select the "best" (least uncertain) segmentation result. For myocardial T1 mapping, image segmentation using a quality control-driven technique has been previously proposed by Hann et al. (22, 23) with impressive results. Their work uses a "label voting" scheme and predicts a Dice similarity score which is in turn used as the measure to select the best segmentation of the T1 map.

Here, we used the U-Net (with space-time patch inputs) as the network architecture for the pool of DNNs in our proposed image analysis framework. U-Net is a well-established and commonly used DNN architecture for segmentation tasks in medical imaging. In CMR image segmentation and clinical reporting, as in other areas where generalizability is a key factor, employing overly complex models that are prone to overfitting are less desirable. Nevertheless, our proposed DAUGS analysis framework (and U-map quantification scheme) is agnostic to the choice of network architecture and can be expanded to include larger model pools with various architectures and complexity levels. Compared to ensemble-based deep learning approaches that derive a single uncertainty map/measure for the pool of DNN models, our sliding-patch U-map approach quantifies the uncertainty for each member of the DNN pool hence enabling a data-adaptive model selection technique for improved robustness to dataset shifts (47, 48).

**Limitations**

In this proof-of-concept work, we used a limited external dataset from two medical centers to demonstrate the potential of the proposed technique. We are working towards evaluating the performance of this approach on larger multi-center datasets. Specifically, the SCMR Registry provides an opportunity for using a large multi-center database of clinical images to test A.I.-powered techniques such as the proposed approach. Our training dataset was limited; however, this is not a fundamental limitation since the technical framework is general and can use larger training datasets to improve the segmentation performance (54). Although the proposed method performed well for the case with noticeable ECG mis-triggering in our test dataset, it may fail in more severe cases of MoCo failure; in such scenarios, alternative techniques such as segmentation of raw free-breathing images may be warranted (25,55). Our internal and external datasets were all acquired at 3T; investigating the effect of lower field strength on the performance of the proposed method will be part of our future work. In terms of the segmentation performance (measured by Dice score in this work), prior work (7) has shown that approaching a Dice score above 0.90 using DNNs may be possible if a sufficiently large training dataset (>10x larger than our study) is used. As has been demonstrated in the field of computational microscopy (49), additional training data can be incorporated by retraining the pool of DNN models to improve the segmentation performance over time. Finally, we developed the model only with a categorical cross-entropy cost function. Comparison of different cost functions remains an open area to explore.

**CONCLUSIONS**

We demonstrate proof-of-concept results for a new data-adaptive deep learning technique that uses an automatically generated uncertainty measure to guide the segmentation process of stress perfusion CMR time series. Our results show that, compared to the established deep learning-based approach, the proposed method significantly improves the fully automatic segmentation performance for external multi-center datasets acquired with a different pulse sequence or scanner vendor. This improved robustness to such dataset shifts has the potential to enable fully automatic reporting of myocardial perfusion MRI studies in the setting of large multi-center datasets.

**Abbreviations**

A.I.: artificial intelligence

AIF: arterial input function

bSSFP: balanced steady-state free precession

CMR: cardiovascular magnetic resonance

DAUGS: Data Adaptive Uncertainty Guided Space-time

DNN: deep neural network

exD: external Dataset

FLASH: fast low angle shot

GRE: gradient-recalled echo

inD: internal Dataset

inD-test: test set portion of the internal dataset

LGE: late gadolinium enhancement

LoA: limit of agreement

LV: left ventricular

MBF: myocardial blood flow

MoCo: motion correction

MRI: magnetic resonance imaging

n.s.: not significant

ROI: region of interest

SR: saturation recovery

U-map: uncertainty map

# DECLARATIONS

**Availability of Data and Software Code**

Although the proposed method is currently an investigational technique and not yet tested in a clinical setting, we have openly provided the source code for the proposed method (DAUGS analysis: patch-level training and generation of U-maps) and the trained models on GitHub: https://github.com/TIM-Lab/DAUGS.

**Competing Interests**

The authors declare that they have no competing interests.

**Funding**

This work was supported by the Lilly Endowment INCITE award to B. Sharif and NIH grant R01-HL153430 (PI: B. Sharif); the Erika Glazer Family Foundation and NIH grants U54-AG065141, R01-HL153500, R01-HL146158, R01-HL148788.

**Authors' contributions**

Data collection, anonymization and analysis (BH, JW, CNBM, RJ, OPS, JWW); initial concept, study design and development of algorithms (DMY, KY, BS); development and implementation of the deep learning techniques (DMY); manual contouring of the images (BH, BS); clinical input about the overall approach (SVR, RD, OPS); interpretation of data and results (DMY, KY, BS); drafting of the manuscript (DMY, BS); critical reviewing of the manuscript (all authors); approval of the submitted version (all authors). All authors read and agreed on the final submitted version of the manuscript.

# REFERENCES

1. Kwong R, Ge Y, Steel K, Bingham S, Abdullah S, Fujikura K, et al. Cardiac Magnetic Resonance Stress Perfusion Imaging for Evaluation of Patients With Chest Pain. J Am Coll Cardiol 2019;74(14):1741–55.

2. Schwitter J, Wacker CM, Wilke N, Al-Saadi N, Sauer E, Huettle K, et al. MR-IMPACT II: Magnetic Resonance Imaging for Myocardial Perfusion Assessment in Coronary artery disease Trial: perfusion-cardiac magnetic resonance vs. single-photon emission computed tomography for the detection of coronary artery disease: a comparative multicentre, multivendor trial for the MR-IMPACT Investigators. Eur Heart J 2013;34(10):775–81.

3. Heitner JF, Kim RJ, Kim HW, Klem I, Shah DJ, Debs D, et al. Prognostic Value of Vasodilator Stress Cardiac Magnetic Resonance Imaging: A Multicenter Study With 48 000 Patient-Years of Follow-up. JAMA Cardiol 2019;4(3):256–64.

4. Hachamovitch R, Pena JM, Xie J, Shaw LJ, JK M. Imaging registries and single-center series. JACC Cardiovasc Imaging 2017;10(3):276-85.

5. Bax JJ, Chandrashekhar Y. The Power of Large Clinical Databases and Registries in our Understanding of Cardiovascular Diseases. JACC Cardiovasc Imaging. 2021;14(11):2272-4.

6. Leiner T, Rueckert D, Suinesiaputra A, Baeßler B, Nezafat R, Išgum I, et al. Machine learning in cardiovascular magnetic resonance: basic concepts and applications. J Cardiovasc Magn Reson 2019;21(1):61.

7. Xue H, Davies RH, Brown LA, Knott KD, Kotecha T, Fontana M, et al. Automated inline analysis of myocardial perfusion MRI with deep learning. Radiology: Artif Intell 2020;2(6):e200009.

8. Scannell CM, Veta M, Villa ADM, Sammut EC, Lee J, Breeuwer M, et al. Deep-Learning-Based Preprocessing for Quantitative Myocardial Perfusion MRI. J Magn Reson Imaging 2020;51(6):1689-96.

9. Gonzales RA, Seemann F, Lamy J, Mojibian H, Atar D, Erlinge D, et al. Mvnet: automated time-resolved tracking of the mitral valve plane in CMR long-axis cine images with residual neural networks: a multi-center, multi-vendor study. J Cardiovasc Magn Reson 2021;23(1):137.

10. Sander J, de Vos BD, Wolterink JM, Išgum I. Towards increased trustworthiness of deep learning segmentation methods on cardiac MRI. Proc SPIE Med Imaging 2019:1094919. DOI: 10.1117/12.2511699

11. Fahmy AS, Neisius U, Chan RH, Rowin EJ, Manning WJ, Maron MS, et al. Three-dimensional Deep Convolutional Neural Networks for Automated Myocardial Scar Quantification in Hypertrophic Cardiomyopathy: A Multicenter Multivendor Study. Radiology 2020;294(1):52-60.

12. Hundley WG, Bluemke DA, Bogaert J, Flamm S, Fontana M. Society for Cardiovascular Magnetic Resonance (SCMR) guidelines for reporting cardiovascular magnetic resonance examinations. J Cardiovasc Magn Reson 2022;24(1):29.

13. Rajpurkar P, MP. L. The Current and Future State of AI Interpretation of Medical Images. N Engl J Med 2023; 388(21):1981-90.

14. Chen C, Qin C, Qiu H, Ouyang C, Wang S, Chen L, et al. Realistic adversarial data augmentation for MR image segmentation. Med Image Comput Comput Assist Interv (MICCAI) 2020:667-77. DOI: 10.1007/978-3-030-59710-8_6

15. Simantiris G, Tziritas G. Cardiac MRI segmentation with a dilated CNN incorporating domain-specific constraints. IEEE J Sel Top Signal Process. 2020;14(6):1235-43. DOI: 10.1007/978-3-030-59710-8_65

16. Zhang M, Levine S, Finn C, editors. Memo: Test time robustness via adaptation and augmentation. Proc Adv Neur Info Process Syst (NeurIPS) 2022: 38629-42. DOI: 10.48550/arXiv.2110.09506

17. Roy AG, Conjeti S, Navab N, Wachinger C, Initiative AsDN. Bayesian QuickNAT: Model uncertainty in deep whole-brain segmentation for structure-wise quality control. NeuroImage 2019;195:11-22.

18. Hoebel K, Andrearczyk V, Beers A, Patel J, Chang K, Depeursinge A, et al. An exploration of uncertainty information for segmentation quality assessment. Proc SPIE Med Imaging 2020:113131K. DOI: 10.1117/12.2548722

19. DeVries T, Taylor GW. Leveraging uncertainty estimates for predicting segmentation quality arXiv:1807.00502, 2018. DOI: 10.48550/arXiv.1807.00502

20. Wickstrøm K, Kampffmeyer M, Jenssen R. Uncertainty and interpretability in convolutional neural networks for semantic segmentation of colorectal polyps. Med Image Analysis 2020;60:101619.

21. Puyol-Antón E, Ruijsink B, Baumgartner CF, Masci P-G, Sinclair M, Konukoglu E, et al. Automated quantification of myocardial tissue characteristics from native T1 mapping using neural networks with uncertainty-based quality-control. J Cardiovasc Magn Reson 2020;22(1):60.

22. Hann E, Popescu IA, Zhang Q, Gonzales RA, Barutçu A, Neubauera S, et al. Deep neural network ensemble for on-the-fly quality control-driven segmentation of cardiac MRI T1 mapping. Med Image Analysis 2021;71:102029.

23. Hann E, Gonzales RA, Popescu IA, Zhang Q, Ferreira VM, Piechnik SK. Ensemble of Deep Convolutional Neural Networks with Monte Carlo Dropout Sampling for Automated Image Segmentation Quality Control and Robust Deep Learning Using Small Datasets. Med Image Underst Analysis 2021:280-93. DOI: 10.1007/978-3-030-80432-9_22

24. Yalcinkaya DM, Youssef K, Heydari B, Zamudio L, Dharmakumar R, Sharif B. Deep Learning-Based Segmentation and Uncertainty Assessment for Automated Analysis of Myocardial Perfusion MRI Datasets Using Patch-Level Training and Advanced Data Augmentation. Proc IEEE Eng Med Biol Soc Conf 2021:4072-78. DOI: 10.1109/EMBC46164.2021.9629581

25. Yalcinkaya DM, Youssef K, Heydari B, Simonetti O, Dharmakumar R, Raman S, Sharif B. Temporal Uncertainty Localization to Enable Human-in-the-loop Analysis of Dynamic Contrast-enhanced Cardiac MRI Datasets. Med Image Comput and Computer-Assisted Interv (MICCAI), 2023:453-462. DOI: 10.1007/978-3-031-43898-1_44

26. Lu MY, Williamson DFK, Chen TY, Chen RJ, Barbieri M, Mahmood F. Data-efficient and weakly supervised computational pathology on whole-slide images. Nat Biomed Eng 2021;5(6):555-70.

27. Kuo W, Häne C, Mukherjee P, Malik J, Yuh EL. Expert-level detection of acute intracranial hemorrhage on head computed tomography using deep learning. Proc Natl Acad Sci USA 2019;116:22737–45. DOI: 10.1073/pnas.1908021116

28. Youssef K, Heydari B, Zamudio L, Beaulieu T, Cheema K, Dharmakumar R, et al. A Patch-Wise Deep Learning Approach for Myocardial Blood Flow Quantification with Robustness to Noise and Nonrigid Motion. Proc IEEE Eng Med Biol Soc Conf 2021:4045-51. DOI: 10.1109/EMBC46164.2021.9629630

29. Yushkevich PA, Piven J, Hazlett HC, Smith RG, Ho S, Gee JC, et al. User-guided 3D active contour segmentation of anatomical structures: significantly improved efficiency and reliability. NeuroImage 2006;31(3):1116-28.

30. Jacobs M, Benovoy M, Chang L-C, Arai AE, Hsu L-Y. Evaluation of an automated method for arterial input function detection for first-pass myocardial perfusion cardiovascular magnetic resonance. J Cardiovasc Magn Reson 2016;18(1).

31. Coupé P, Manjón JV, Fonov V, Pruessner J, Robles M, Collins DL. Patch-based segmentation using expert priors: Application to hippocampus and ventricle segmentation. NeuroImage 2011;54:940-54.

32. Huo Y, Xu Z, Xiong Y, Aboud K, Parvathaneni P, Bao S, et al. 3D whole brain segmentation using spatially localized atlas network tiles. NeuroImage 2019;194:105-19.

33. Vu QD, Graham S, Kurc T, To MNN, Shaban M, Qaiser T, et al. Methods for Segmentation and Classification of Digital Microscopy Tissue Images. Front Bioeng Biotechnol 2019;7(53).

34. Lee B, Yamanakkanavar N, Choi JY. Automatic segmentation of brain MRI using a novel patch-wise U-net deep architecture. PLoS One 2020;15(8):e0236493.

35. Li H, Chen M, Wang J, Illapani VSP, Parikh NA, He L. Automatic Segmentation of Diffuse White Matter Abnormality on T2-weighted Brain MR Images Using Deep Learning in Very Preterm Infants. Radiology: Artif Intell 2021;3(3):e200166.

36. Rudie JD, Weiss DA, Colby JB, Rauschecker AM, Laguna B, Braunstein S, et al. Three-dimensional U-Net Convolutional Neural Network for Detection and Segmentation of Intracranial Metastases. Radiology: Artif Intell 2021;3(3):e200204.

37. Ronneberger O, Fischer P, Brox T. U-Net: Convolutional Networks for Biomedical Image Segmentation. Med Image Comput Comput Assist Interv (MICCAI) 2015:234-41. DOI: 10.1007/978-3-319-24574-4_28

38. Eijgelaar RS, Visser M, Müller DMJ, Barkhof F, Vrenken H, Herk Mv, et al. Robust Deep Learning–based Segmentation of Glioblastoma on Routine Clinical MRI Scans Using Sparsified Training. Radiology: Artif Intell 2020;2(5):e190103.

39. Schmidt AM, Desai AD, Watkins LE, Crowder HA, Black MS, Mazzoli V, et al. Generalizability of deep learning segmentation algorithms for automated assessment of cartilage morphology and MRI relaxometry. J Magn Reson Imaging 2023;57(4):1029-39.

40. Rauschecker AM, Gleason TJ, Nedelec P, Duong MT, Weiss DA, Calabrese E, et al. Interinstitutional portability of a deep learning brain MRI lesion segmentation algorithm. Radiology: Artif Intell 2021;4(1):e200152.

41. Lakshminarayanan B, Pritzel A, Blundell C. Simple and Scalable Predictive Uncertainty Estimation using Deep Ensembles. Proc Adv Neur Info Process Syst (NeurIPS) 2017:6405–16. DOI: 10.48550/arXiv.1612.01474

42. Wenzel F, Snoek J, Tran D, R. J. Hyperparameter Ensembles for Robustness and Uncertainty Quantification. Proc Adv Neur Info Process Syst (NeurIPS) 2020:6514–27, Article No: 546. DOI: 10.48550/arXiv.2006.13570

43. Fort S, Hu H, Lakshminarayanan B, editors. Deep Ensembles: A Loss Landscape Perspective. Proc Bayes Deep Learn Workshop at NeurIPS; 2019. DOI: 10.48550/arXiv.1912.02757

44. Cao Y, Geddes TA, Yang J, P. Y. Ensemble deep learning in bioinformatics. Nat Mach Intell 2020;2:500–8.

45. Ovadia Y, Fertig E, Ren J, Nado Z, Sculley D, Nowozin S, et al. Can You Trust Your Model's Uncertainty? Evaluating Predictive Uncertainty Under Dataset Shift. Proc Neur Info Process Syst (NeurIPS) 2019: 14003–14, Article No: 1254. DOI: 10.48550/arXiv.1906.02530

46. Tiu E, Talius E, Patel P. Expert-level detection of pathologies from unannotated chest X-ray images via self-supervised learning. Nat Biomed Eng 2022;6:1399–406.

47. Liang W, Tadesse GA, Ho D, Fei-Fei L, Zaharia M, Zhang C, et al. Advances, challenges and opportunities in creating data for trustworthy AI. Nat Mach Intell 2022;4(8):669-77

48. Abe T, Buchanan EK, Pleiss G, Zemel R, JP C. Deep Ensembles Work, But Are They Necessary? Proc Adv Neur Info Process Syst (NeurIPS) 2022. DOI: 10.48550/arXiv.2202.0698

49. Stringer C, Wang T, Michaelos M, Pachitariu M. Cellpose: a generalist algorithm for cellular segmentation. Nat Methods 2021;18:100-6.

50. He K, Zhang X, Ren S, J. S. Delving deep into rectifiers: Surpassing human-level performance on imagenet classification. Proc IEEE Int Conf Comp Vis 2015:1026-34. DOI: 10.1109/ICCV.2015.123

51. Jacobs M, Benovoy M, Chang LC, Corcoran D, Berry C, Arai AE, et al. Automated segmental analysis of fully quantitative myocardial blood flow maps by first-pass perfusion cardiovascular magnetic resonance. IEEE Access. 2021;9:52796-811

52. Jerosch-Herold M. Quantification of myocardial perfusion by cardiovascular magnetic resonance. J Cardiovasc Magn Reson. 2010;12(1):1-6.

53. Ng M, Guo F, Biswas L, Petersen SE, Piechnik SK, Neubauer S, Wright G. Estimating uncertainty in neural networks for cardiac MRI segmentation: a benchmark study. IEEE Trans Biomed Eng. 2022;70(6):1955-66.

54. Yalcinkaya DM, Li Z, Youssef K, Zamudio L, Polsani V, Elliott E, et al. Automatic Segmentation of Multi-center Multi-field-strength Perfusion CMR Datasets with Deep Learning-based Uncertainty-guided Analysis: Preliminary Findings Using the SCMR Registry, Proc SCMR, J Cardiovasc Magn Reson. DOI: 10.1016/j.jocmr.2024.100964

55. Yalcinkaya DM, Youssef K, Li Z, Heydari B, Dharmakumar R, Judd R, et al. Clinician-in-the-loop Analysis of Free-Breathing Stress Perfusion CMR Datasets with Dynamic Quality Control: Preliminary Evaluation Using the SCMR Registry, Proc SCMR, J Cardiovasc Magn Reson. DOI: 10.1016/j.jocmr.2024.100990

# SUPPLEMENTARY MATERIAL: APPENDICES

**Appendix A – Additional Details for the Data Augmentation and Model Training Processes**

Data augmentation was performed to boost the training set by ≈40-fold using random rotations (±60°), shear (±10°), translations (±2 pixels), uniform scaling (randomly selected scaling factor in the [0.8, 1.2] range), flat-field correction (using the "imflatfield" routine in Matlab, Mathworks, Natick, MA) with 50% probability ($\sigma \in [0, 5]$), and gamma correction with 50% probability ($\gamma \in [0.5, 1.5]$) applied to each perfusion image series in the training dataset. The augmented training dataset contained 137,160 space-time patches.

For training of the DNN models in this study, we used a cross-entropy loss with the Adam optimizer and initialized the network parameters using the He initializer (50) and employed a batch size of 128. A linear learning rate drop was utilized at every two epochs with a drop factor of 0.5 with an initial learning rate of $5 \times 10^{-4}$. We ran the training process for a maximum of 15 epochs and employed "early stopping" if the myocardial Dice score on the validation set was not improving over 5 epochs. It should be noted that, compared to image-level models, patch-level models tend to converge in fewer epochs thanks to the multi-fold increase in the size of the training dataset.

**Appendix B – Additional Details for Patch-level DNN Training and Analysis**

To further optimize the patch-level analysis technique, we experimented with different patch sizes and the spatial sliding-window stride. Among a set of three candidate patch sizes (16×16, 32×32, 64×64), a spatial patch size of 64×64 resulted in the best performance (validation dataset). The networks trained with smaller patch sizes (16×16 or 32×32) had difficulty focusing on global features (e.g., position of the septal wall relative to the location of bloodpool, etc.), resulting in inferior performance. As a result of working with a relatively large patch size 64×64, nearly all extracted patches included features from the heart ROI. Next, having selected a patch size of 64×64, we evaluated the effect of stride size in a similar fashion. We found that a relatively large stride size of 50% (i.e., 32 pixels with a patch size of 64×64) performed well. Finally, we used a much smaller stride size of 3% (2 pixels) for computing the U-maps. Using this small stride size helps to minimize the "edge effects" and potentially allows for a more accurate pixel-wise U-map since it enables a single spatiotemporal pixel location to be analyzed (segmented) many times.

As described in Fig. 3(a), patch-level segmentation results need to be combined to generate the final (full ROI) analysis. This process involves a "voting method" whereby each pixel in the final segmentation result is classified as being inside/outside of the LV myocardium (class 0 and class 1, respectively). We experimented with two voting methods to decide the class of each pixel when reconstructing full-ROI images from patches: (i) majority voting wherein a pixel is determined to belong to class 1 if at least half of the patches that it belongs to are showing a probability >0.5 for it to be a myocardial pixel, (ii) computing the mean of the softmax probabilities for a given pixel and applying a binarization (0.5 threshold). The results in the validation dataset did not exhibit a significant difference in terms of segmentation performance (Dice score); therefore, we proceeded with the mean of the softmax probabilities.

**Appendix C – Further details on $U_{pp}$ calculation and derivation of the U-map range**

First, we provide a mathematic description of the patch combination process that is pictorially described in Fig 3(a). The binary segmentation solution, denoted by S-map, at a given spatial coordinate $(x, y)$ is derived from the mean of the probability scores from the patches that are in $\Gamma_{(x,y)}$ followed by a binarization operation, that is:

$$\text{S-map}(x, y) = \begin{cases} 1, & if \ \frac{1}{|\Gamma_{(x,y)}|}\sum_{i=1}^{|\Gamma_{(x,y)}|} p^i \geq 0.5 \\ 0, & otherwise. \end{cases}$$

It is also worth mentioning that $N_{myo}$, the number of myocardial pixels in the segmentation solution, can be derived as:

$$N_{myo} = \sum_{x,y} \text{S-map}(x, y)$$

Additionally, the maximum value that a pixel can reach in the U-map is achieved when there is a maximum level of variation in the probability scores $p^i$ corresponding to that pixel location. This "maximum level of discrepancy" between patches for a fixed pixel location $(x, y)$ happens when half of the patches have a probability score of zero for this pixel location and the other half have a probability score of one at the same pixel location (which also implies that the mean value across all $\Gamma_{(x,y)}$ patches at $(x, y)$ will be 0.5). Hence, the standard deviation of these softmax probabilities, which corresponds to the maximum value U-map$(x, y)$ can assume, is:

$$\text{SD}_{\max} = \sqrt{\frac{\sum_i |p^i - 0.5|^2}{|\Gamma_{(x,y)}|}} = \sqrt{\frac{\frac{|\Gamma_{(x,y)}|}{2}(1-0.5)^2 + \frac{|\Gamma_{(x,y)}|}{2}(0-0.5)^2}{|\Gamma_{(x,y)}|}} = 0.5$$

On the other hand, U-map$(x, y)$ is minimized when there is consensus regarding the rounded softmax probability score at $(x, y)$ among all the patches that this pixel belongs to, which would result in a standard deviation of $\text{SD}_{\min} = 0$. Therefore, the dynamic range of the U-map is in the [0, 0.5] range.

**Appendix D - Effect of Simulated Motion-correction Error on the Uncertainty Metric**

To complement our prior work on verifying the utility of U-map in interpreting patch-level segmentation of CMR perfusion datasets (24, 25), we conducted a set of simulation experiments to demonstrate that our mean per-pixel uncertainty metric ($U_{pp}$) is able to track the level uncertainty associated with difficulty/challenge in myocardial segmentation by varying the quality of nonrigid motion correction (MoCo) in an example stress image series. In general, the more MoCo errors we have in a test case, the more uncertain we expect the DNN-derived segmentation to be. Our simulations results, presented in the Suppl. Figure S1, are consistent with this observation (details provided in the caption).

**Appendix E – Impact of improved segmentation accuracy on fully quantitative myocardial blood flow analysis**

We evaluated the impact of automatic DNN-derived segmentation accuracy on fully quantitative analysis of stress perfusion CMR in two of the three test sets: inD-test and exD-1. Specifically, we compared the segment-wise MBF values obtained based on the segmentation contours from the proposed DAUGS analysis and the established approach using the Fermi-constrained deconvolution for dual-bolus stress perfusion CMR (51, 52). Landmark (RV insertion point) detection and the division of the segmentation solutions into 6-segment was automatically performed thanks to the RV segmentation ability of all trained DNNs in this work. Supplementary Fig. S2 shows the MBF results for inD-test which shows that proposed DAUGS analysis approach performed similarly to the established DNN-based analysis ($R^2$=0.91, p<0.0001 for both). However, on exD-1, the MBF values corresponding to the segmentations obtained from the proposed approach resulted in a stronger correlation compared to the established approach with fewer outliers ($R^2$=0.95 vs. $R^2$=0.87) and tighter 95% limits of agreement in Bland-Altman analysis (details provided in Suppl. Figure S3). These results parallel the segmentation performance gap between two methods in terms of Dice score. In Suppl. Figure S3(c), we highlight three outliers from panel (a), which correspond to 3 myocardial segments belonging to the same perfusion image series. As can be seen from (b) and (c), in contrast to the established approach, the proposed method provides a relatively accurate segmentation which is reflected in how the MBF numbers for these 3 segments (labeled as A, B, and C) are distributed in the two plots in (b).

**Appendix F – Heterogeneity of segmentation solutions in the model pool for the two external datasets**

In the proposed DAUGS analysis approach, the fifty-member model pool (Fig. 1) is created by including the DNN models obtained during the training process (on the internal dataset) from different training sessions/runs (each with a different parameter initializations), or by including the intermediate checkpoints. Here, we examined the heterogeneity (diversity) of the segmentation solutions in the DNN model pool for the two external datasets by visualizing the segmentation solutions across all members of the model pool, the corresponding U-maps, and the histogram of the uncertainty metric ($U_{pp}$ values). Specifically, we picked two external patient studies: one from exD-1 and one from exD-2.

The results are shown in Suppl. Figure S4 (caption includes additional details). As can be seen from the figure, in both examined cases, there is a high level of heterogeneity among the model pool members in terms of quality of segmentation (including failed or non-contiguous segmentation), the U-map composition, and the corresponding $U_{pp}$ metric when there is a need to deal with "dataset shifts" that are present in the external datasets. Furthermore, there is not a particular model in the model pool that does best for all samples of the external dataset (also observed in Suppl. Fig. S4) which is consistent with the DAUGS analysis approach wherein the model selection process is data-adaptive (bottom half of Fig. 1).

**Appendix G – Alternatives to $U_{pp}$ that can be derived from U-map**

The data-adaptive model selection step in the proposed DAUGS analysis technique is based on a metric of uncertainty, which is used to rank the segmentation solutions of the model pool and pick the one with the lowest uncertainty as the "best" solution. In Methods and Results, we defined and used the mean "per pixel" uncertainty metric, $U_{pp}$, for this purpose. However, there are many potential alternatives to $U_{pp}$ and exploring the "optimal" design of a mapping from the two-dimensional U-map to a scalar uncertainty metric remains as future work. Here we explored a potential alternative to $U_{pp}$, that is, the total energy of the U-map (without normalization by the number of myocardial pixels): $U_{tot} = ||U - map||_F^2$.

We then repeated the experiments (those described in the first subsection of Results) to evaluate the segmentation performance using this alternative uncertainty metric ($U_{tot}$ instead of $U_{pp}$) for all three test sets (inD-test, exD-1, and exD-2). The resulting Dice scores were: 0.896 ± 0.048 for inD-test, and 0.878 ± 0.053 for exD-1, and 0.797 ± 0.142 for exD-2. These average Dice scores are slightly lower than those obtained with $U_{pp}$ (Fig. 4) although the difference is not statistically significant for any of the test sets. By inspecting the case by case differences (in exD-1 and exD-2), we noticed that, in challenging cases (e.g., thin apical slices) the $U_{tot}$-selected solutions tend to have more noncontiguous contours. An example from exD-1 is shown in Suppl. Figure S5. This improved performance can be explained by the fact that $U_{pp}$ (unlike $U_{tot}$) can (to some extent) regularize the model selection process (which can be thought of as an "inverse problem" with side information) by penalizing the solutions that have very few myocardial pixels. This type of regularization in effect encodes a form of "spatial information" from the U-map into the scalar metric $U_{pp}$. It remains to be seen if an alternative "functional" (a general form of mapping from the 2D U-map to a scalar) that is more sophisticated than $U_{pp}$ can be designed to further improve the segmentation performance.

**Appendix H – Additional evaluation of the generalization capability of the proposed method**

To further evaluate the generalization capability of the proposed DAUGS analysis method, i.e., its performance on the external datasets, we compared its performance to a new DNN that does not suffer from "dataset shifts" (13). To this end, we trained this dataset-shift-free DNN with partitioned the entire external data (combining exD-1 and exD-2) into training, validation, and testing subsets as described below:

— external training subset: n = 40 (n = 30 from exD-1; n = 10 from exD-2);

— external validation subset: n = 4 (n = 3 from exD-1; n = 1 from exD-2);

— external test subset: n = 11 (n = 7 from exD-1; n = 4 from exD-2).

During training the same augmentation setting described in Appendix A were applied and the external training subset (n = 30 from exD-1; n = 10 from exD-2) was augmented 40-fold. Similarly to the established method, we ran 5 independent training sessions and the best performing DNN was chosen based on the segmentation performance (Dice) on the external validation subset (n = 3 from exD-1; n = 1 from exD-2). This dataset-shift-free DNN resulted in a Dice score of 0.897 ± 0.10 on the external test subset (n = 7 from exD-1; n = 4 from exD-2). On this same external test subset, the network trained on the internal dataset (with no access to the external datasets) using the DAUGS analysis approach achieved a Dice score of 0.867 ± 0.06. Although the mean Dice score for the latter network is slightly lower than the dataset-shift-free DNN (as expected) the difference was not statistically significant (p=0.3).

# TABLES

**Table 1. Summary of the characteristics of the internal dataset and the two external datasets.**

|  | Internal Dataset (inD) | | | External Dataset 1 (exD-1) | External Dataset 2 (exD-2) |
|---|---|---|---|---|---|
|  | **Training Set** | **Validation Set** | **Internal Test Set** | | |
| **# of subjects** | n = 75 | n = 10 | n = 10 | n = 40 | n = 15 |
| **Scanner** | 3T Siemens Verio | 3T Siemens Verio | 3T Siemens Verio | 3T Siemens Skyra | 3T GE Discovery MR750w |
| **Pulse sequence** | SR-prepared GRE | SR-prepared GRE | SR-prepared GRE | SR-prepared bSSFP | SR-prepared GRE |
| **Study date** | 10/2015 – 10/2017 | 10/2015 – 10/2017 | 12/2017 – 10/2019 | 03/2016 – 02/2018 | 01/2015 – 09/2017 |
| **No. of females** | 68 (91%) | 9 (90%) | 9 (90%) | 10 (25%) | 1 (7%) |
| **Age (years)** | 56.8 ± 11.8 | 55.8 ± 10.7 | 62.7 ± 7.1 | 60.1 ± 14.3 | 66.0 ± 9.8 |
| **BMI (kg/m$^2$)** | 27.0 ± 5.4 | 25.7 ± 4.7 | 32.0 ± 9.8 | 30.3 ± 5.6 | 29 ± 5.28 |
| **Healthy controls (%)** | 12% | 10% | 0% | 0% | 0% |
| **Prevalence of ischemia** | 41 (54.7%) | 5 (50%) | 3 (30%) | 17 (42.5%) | 14 (93%) |
| **Prevalence of positive LGE** | 3 (4%) | 0 (0%) | 0 (0%) | 20 (50%) | 13 (86.7%) |

**Table 1. Summary of the characteristics of the internal dataset (inD) and the two external datasets (exD-1 and exD-2).** There were notable differences between the internal dataset and external datasets in terms of scanner vendor (Siemens vs. GE), subject characteristics, data acquisition protocol (choice of SR-prepared spoiled GRE vs. SR-prepared bSSFP) and pulse sequence parameters (SR time, resolution, flip angle, etc.). Training of the DNNs used a subset of the internal dataset and there was no overlap between the training data and testing data. **Pulse sequence for inD**: SR-prepared FLASH (RF-spoiled GRE) sequence; SR time: 95-105 msec; flip angle: 12; image matrix: 224x160; TGRAPPA rate: 3-4; TR/TE: 2.5/0.9 ms; in-plane resolution $\simeq 1.8 \times 1.8$ mm$^2$; slice thickness: 8 mm. **Pulse sequence for exD-1**: SR-prepared bSSFP sequence; SR time: 95-130 msec; flip angle: 26-35; image matrix: (192-224)x(120-160); TGRAPPA rate: 2; TR/TE: (2.5-2.6)/(1.0-1.1) ms; in-plane resolution $\simeq 1.9 \times 1.9$ mm$^2$; slice thickness: 8 mm. **Pulse sequence for exD-2**: SR-prepared fast (RF-spoiled) GRE sequence; SR time: 119-138 msec; flip angle: 20; matrix: 160x160; ASSET factor 2; TR/TE: 2.5/0.9 ms; in-plane resolution $\simeq 1.7 \times 1.7$ mm$^2$; slice thickness: 10 mm.

# FIGURES

## Figure 1 (Central Illustration)

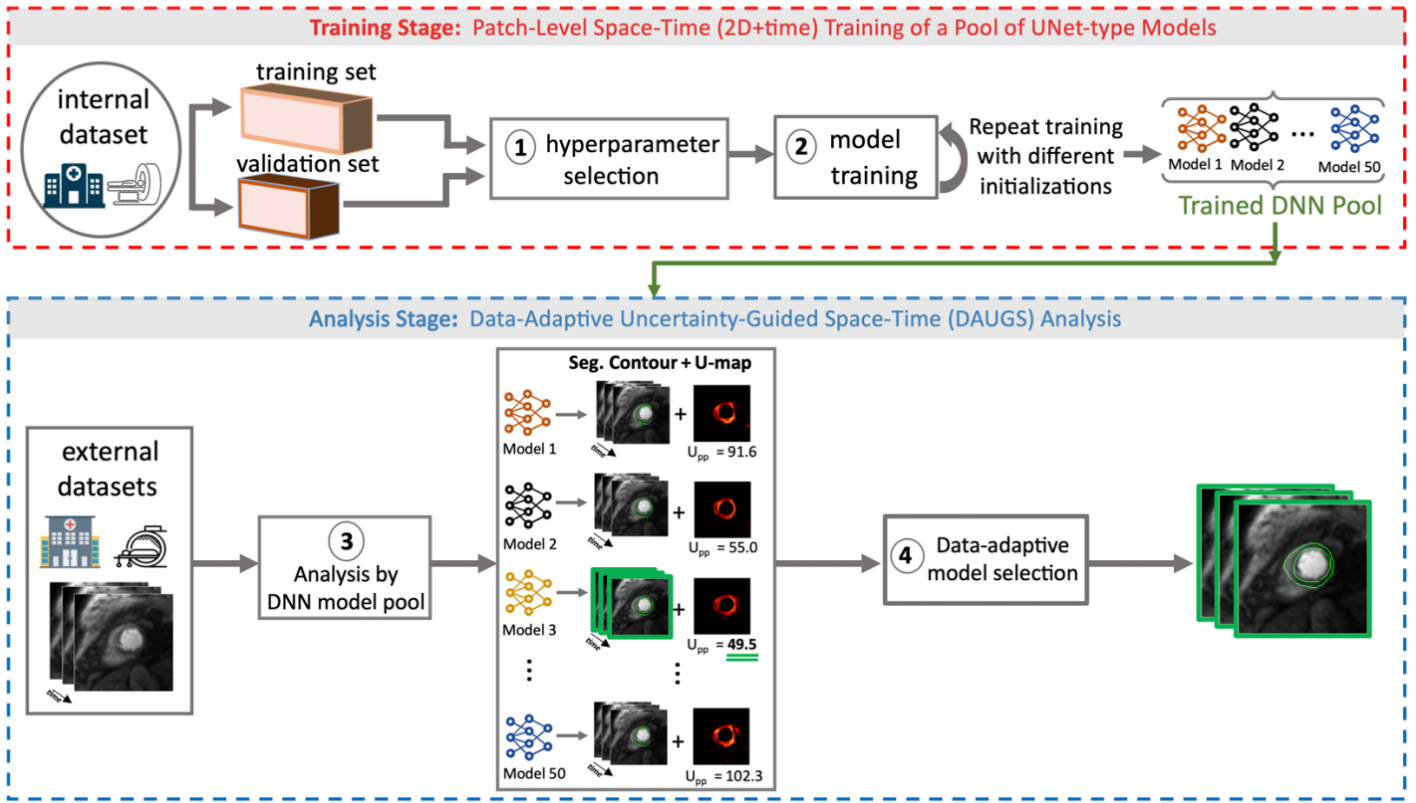

**Figure 1 (Central Illustration).** Description of the proposed data-adaptive uncertainty-guided space-time (DAUGS) analysis approach, which aims to improve the robustness of deep learning-based segmentation of stress perfusion images to variations in software (pulse sequence) and hardware (scanner vendor) in external multi-center datasets. **Training Stage – Steps 1 and 2**: Using a patch-level approach with a fixed "vanilla" spatiotemporal (2D+time) U-Net architecture (24, 25) and a fixed set of hyperparameters, a pool of fifty deep neural network (DNN) models was obtained by running the training process five times using the internal dataset, each time with a different set of initializations for the network parameters (weights), and including intermediate models obtained during the training runs in the model pool (10 DNN models obtained from each of the 5 training runs). **Analysis Stage – Step 3:** In the analysis stage, *each model* in the DNN pool provides a segmentation solution (endo/epi contours) and a corresponding pixel-wise uncertainty map ("U-map") as a byproduct of the segmentation process. **Analysis Stage – Step 4:** Our proposed data-adaptive model-selection approach chooses the segmentation solution with the lowest mean per-pixel energy in its U-map (denoted by $U_{pp}$ in the figure) as the "best" segmentation result.

**Figure 2**

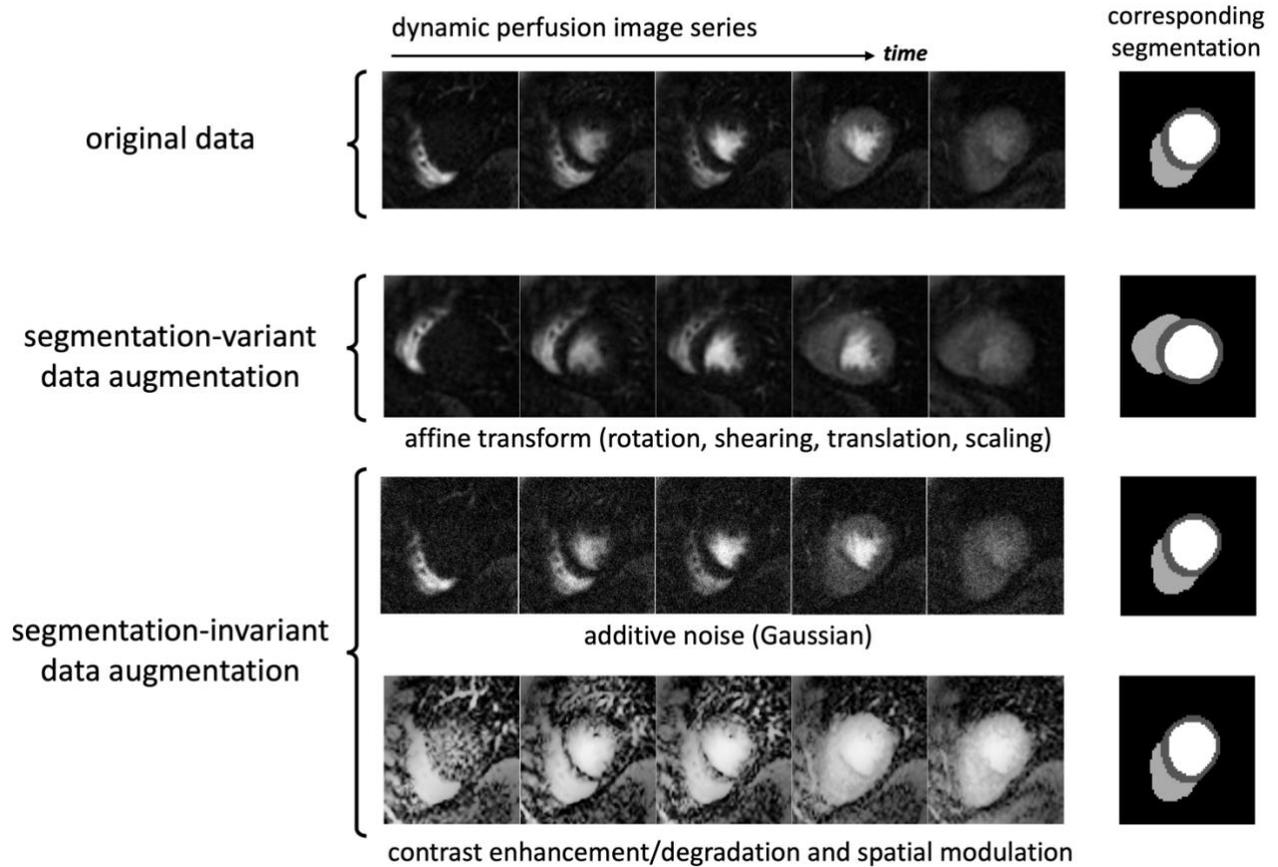

**Figure 2. Data augmentation techniques applied during the DNN training process. (a)** The first type used affine transformation (rotation, shearing, translation, scaling) which is a form of segmentation-variant augmentation, i.e., it involves transforming the manual (original) contours during the augmentation process to generate the corresponding ground-truth segmentation. **(b)** The second type of data augmentation used additive noise, contrast enhancement/degradation and spatial modulation as previously proposed (24), which are segmentation-invariant, that is, they do not affect the location or shape of the corresponding ground-truth contours.

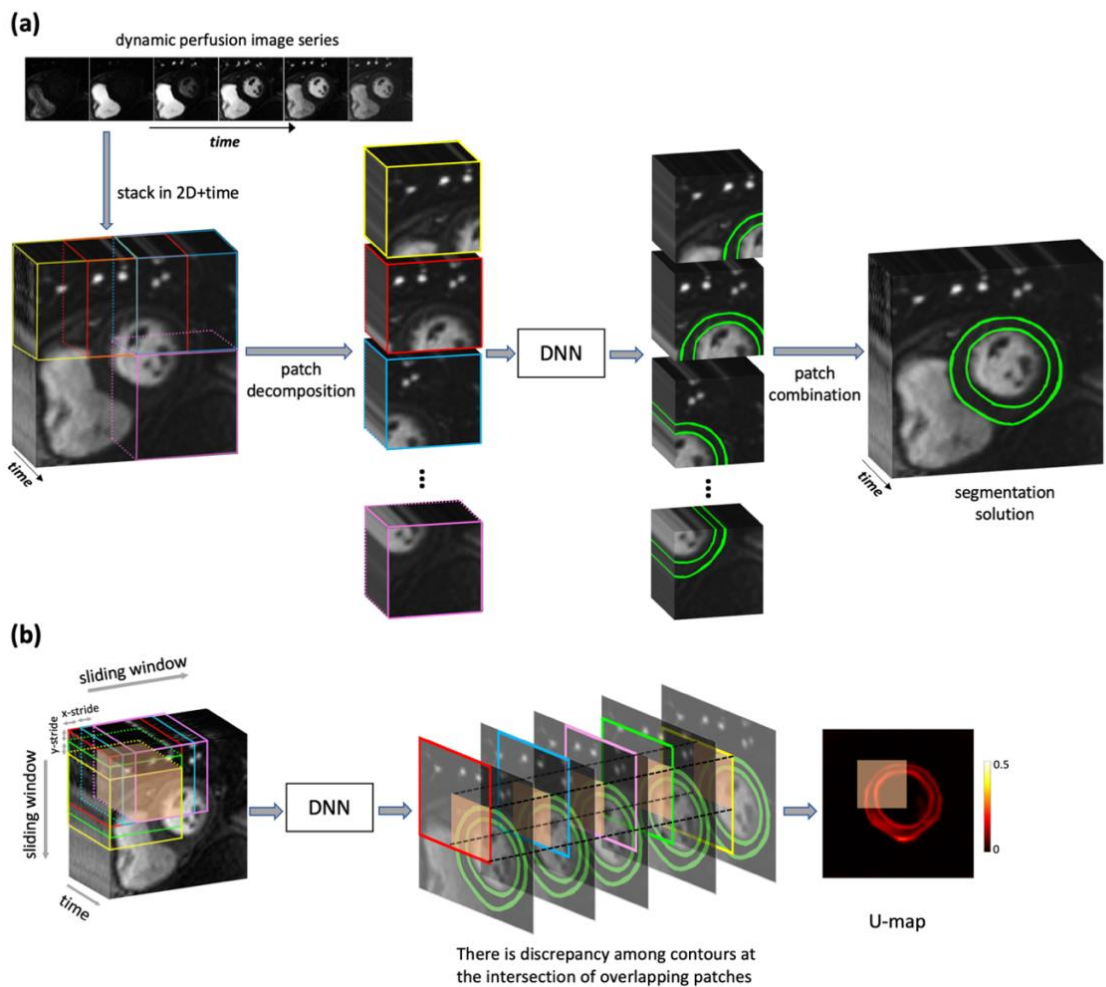

**Figure 3. Methodology for (a) patch-level segmentation of a dynamic perfusion image series using a 2D+time DNN, and (b) pixel-wise mapping of uncertainty for the resulting segmentation solution (corresponding to the same DNN) in the analysis stage. (a)** The data processing pipeline for the proposed patch-level approach that segments dynamic (2D+time) myocardial perfusion image series. First, the motion-corrected perfusion image series is decomposed into space-time patches by applying a spatially sliding window. The decomposed patches are then fed to a DNN with multi-channel U-Net architecture, which jointly processes time frames for each patch to detect the myocardial pixels. The segmented space-time patches (outputs of the DNN) are combined to yield the segmentation solution. **(b)** With a small step size (stride) for the patch-level sliding-window, each pixel in the perfusion image series belongs to multiple patches. For example, the orange-shaded volume shown in the left side of panel (b), lie at the intersection of five overlapping space-time patches and therefore is segmented multiple times during the analysis stage. The discrepancy between the DNN-derived segmentation of these overlapping patches is used to compute a pixel-wise uncertainty map (U-map) as shown here.

**Figure 4**

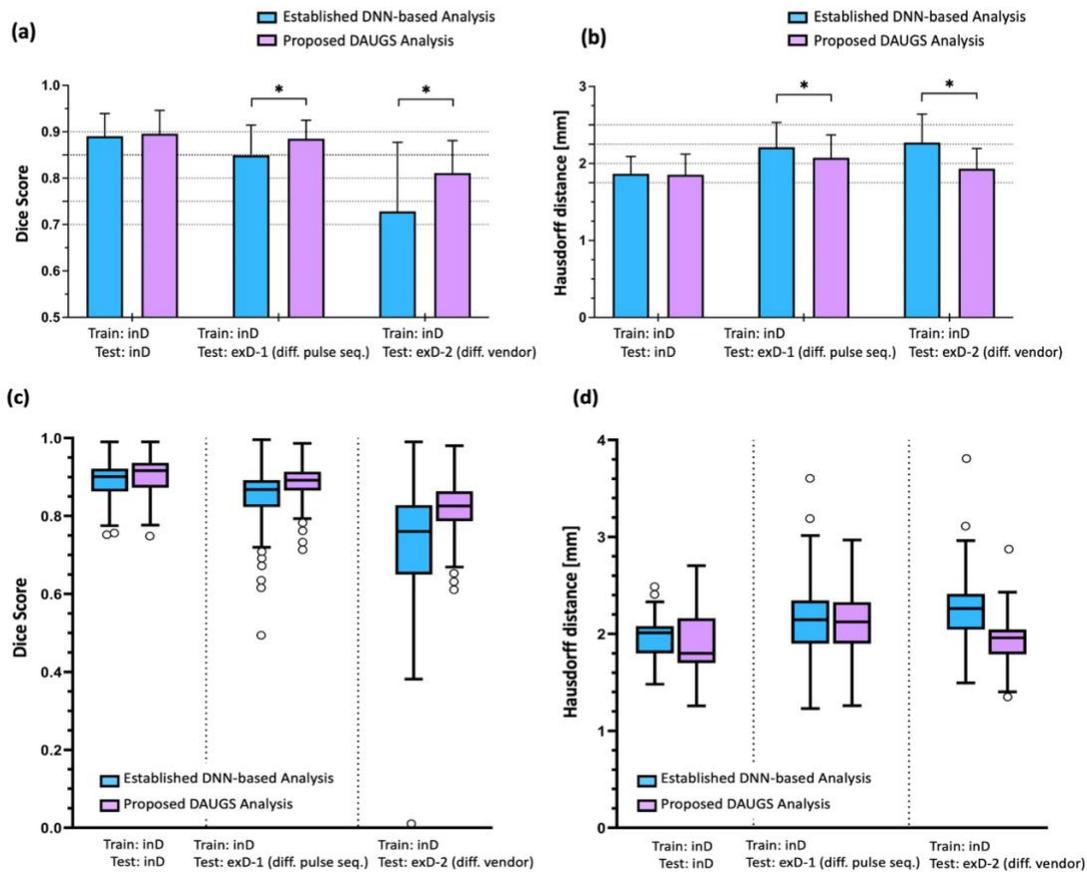

**Figure 4. Segmentation performance of the proposed DAUGS analysis approach vs. the established approach for DNN-based analysis.** The "established" DNN-based analysis refers to the typical DNN training approach in which, the DNN that achieved the highest segmentation accuracy (Dice score) on the validation dataset (among the multiple trained DNNs in the model pool) is selected and used to analyze the test datasets. **(a,c)** Comparison of mean Dice score across the three test datasets (inD-test, exD-1, exD-2) for the proposed approach vs. the established approach. For the internal test set (SR-prepared FLASH; Siemens), the proposed approach performed slightly better, although this outperformance was not significant (p = n.s.). In contrast, on the exD-1 (SR-prepared bSSFP Siemens) and exD-2 (SR-prepared GRE; GE Healthcare), the proposed method significantly outperformed the established approach (p < 0.005). This shows improved generalization ability for the proposed DAUGS analysis approach, which in turn implies improved robustness to variations in the pulse sequence (FLASH vs. SR-bSSFP) or scanner vendor (Siemens vs. GE) compared to the established approach. **(b,d)** Comparison of cumulative Hausdorff distance across the three test datasets for the proposed approach compared to the established approach, which is consistent with the Dice score analysis in (a,b).

**Figure 5**

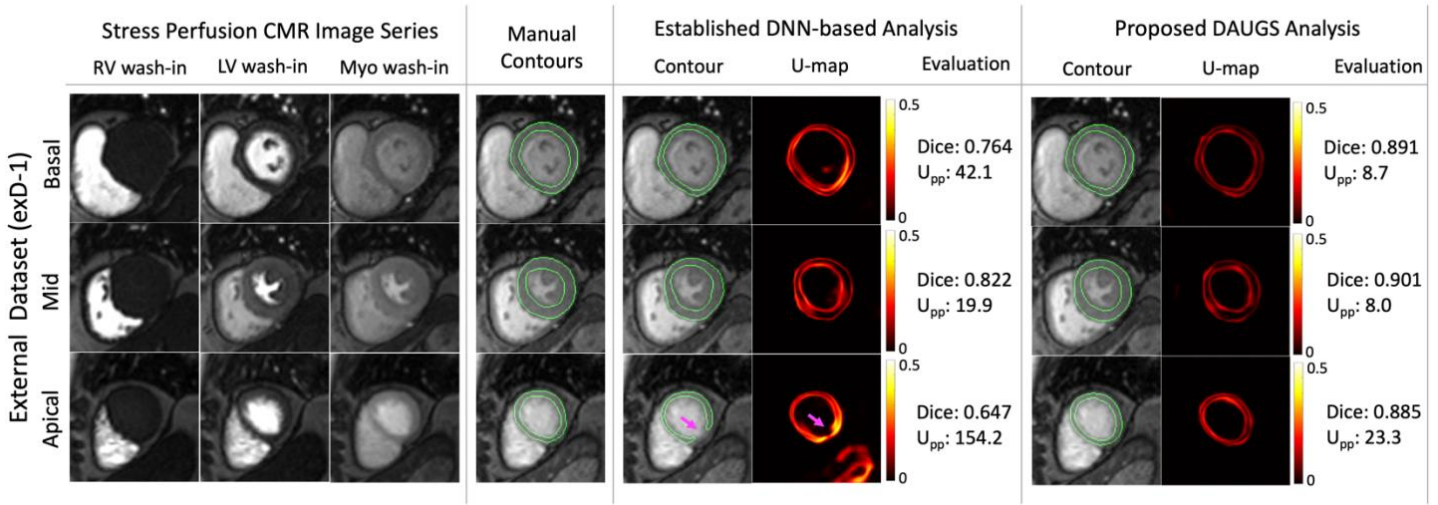

**Figure 5. Representative segmentation results for a patient from exD-1 dataset with normal stress perfusion exam.** Vasodilator stress perfusion CMR image series from the external dataset are shown at three stages of the contrast agent wash-in (right ventricular bloodpool enhancement, LV bloodpool enhancement, and myocardial enhancement) for three short-axis myocardial slices (basal, mid, and apical) together with manual (ground truth) contours and DNN-derived automatic segmentation comparing the established DNN-based analysis vs. the proposed DAUGS analysis approaches. For both approaches, the pixel-wise uncertainty map (U-map) is also shown. For this representative case, the proposed approach accurately delineates the endo/epi contours for all three slices whereas the established approach results in a noncontiguous (failed) segmentation for the apical slice (highlighted by the arrow). Importantly, this error is also reflected in the corresponding U-map as well as $U_{pp}$ which quantifies the normalized per-pixel uncertainty.

**Figure 6**

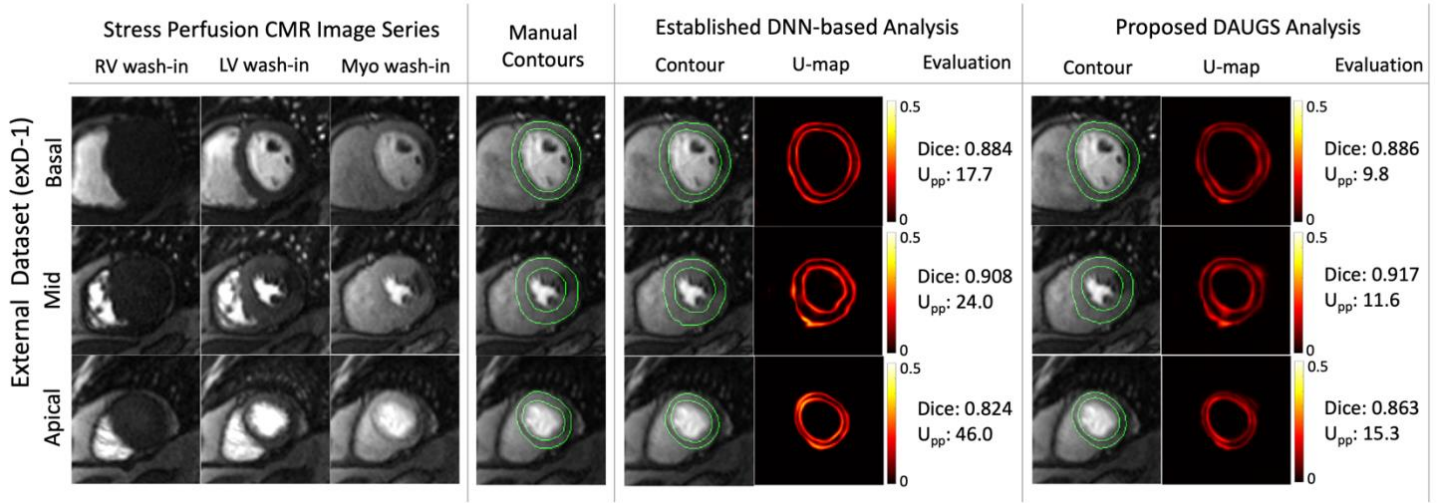

**Figure 6. Representative segmentation results for a patient from exD-1 dataset with stress-induced perfusion defect.** The ground truth and DNN-derived automatic segmentation results are shown (in the same format as Fig 5) comparing the established vs. proposed approach. While both approaches performed well for this case, the proposed DAUGS analysis approach shows a lower level of uncertainty based on the corresponding U-maps and its normalized per-pixel energy ($U_{pp}$).

**Figure 7**

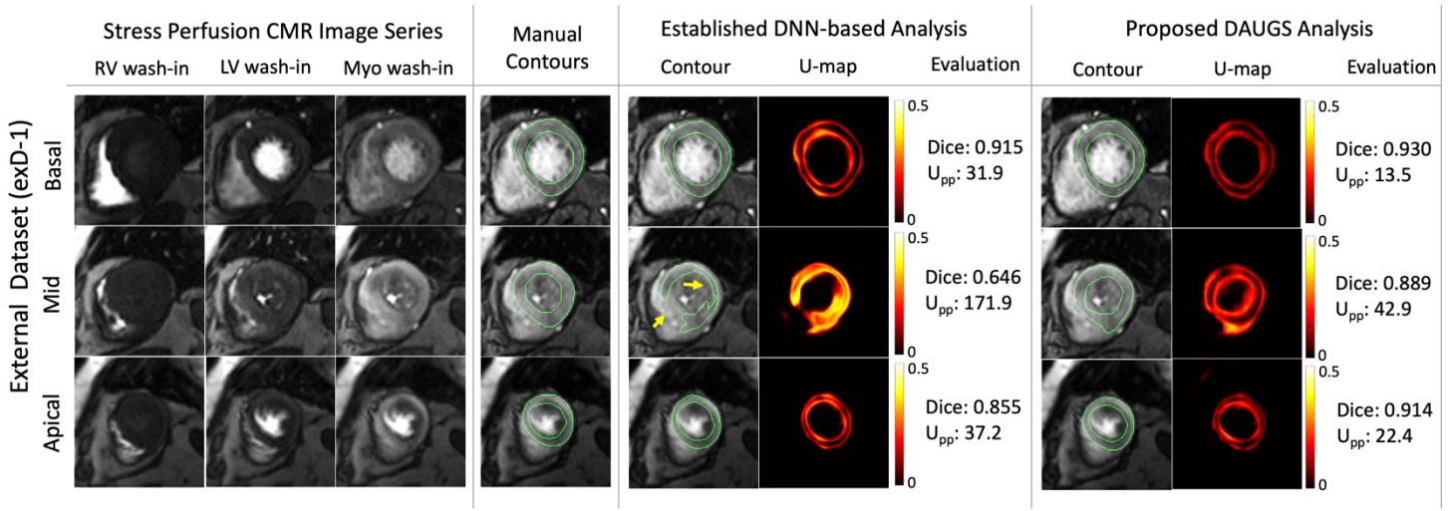

**Figure 7. Representative segmentation results for a challenging case from exD-1 dataset.** Vasodilator stress perfusion images for a patient with hypertrophy and diffuse stress-induced ischemia in all three short-axis slices. Notably, the mid slice, which is acquired at end systole, is challenging to segment due to the difficulty in delineating the subendocardial/subepicardial borders especially in the septal region. This is reflected in the segmentation result for the established approach which fails to segment the mid slice (highlighted by yellow arrows). The proposed DAUGS analysis approach, however, performs well with a mean Dice score of >0.90 across the three slices (Dice: 0.89 for the mid slice with a small error in the infero-septal epicardial contour). The challenging nature of the segmentation task for the mid slice is reflected in the corresponding U-maps and the normalized per-pixel uncertainty ($U_{pp}$) for both the established and proposed approaches.

**Figure 8**

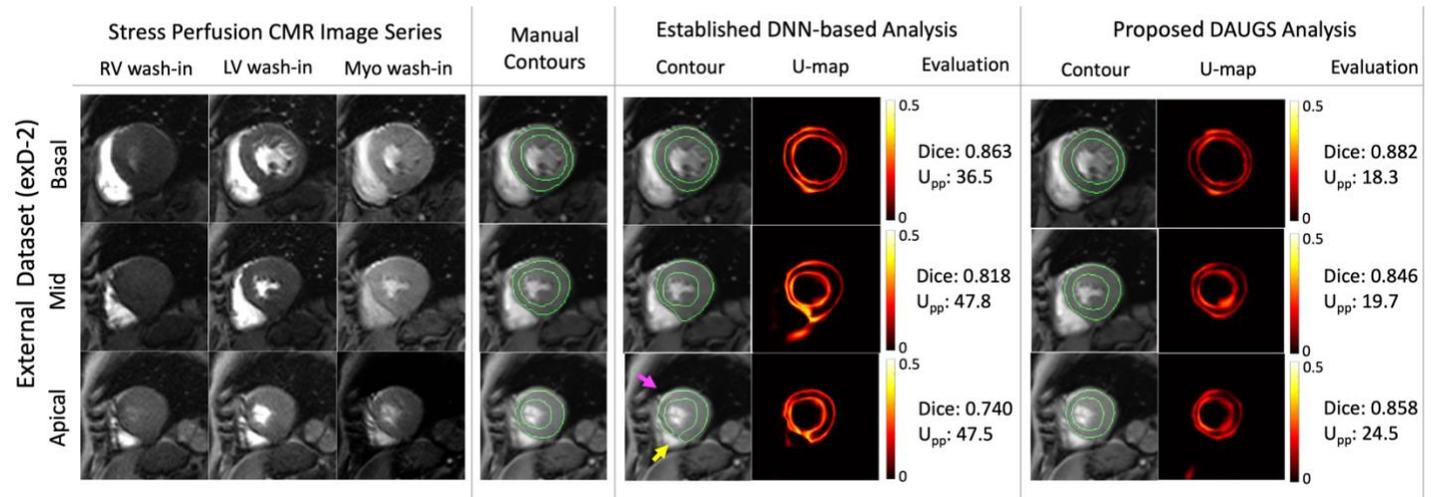

**Figure 8. Representative segmentation results for a patient from exD-2.** Stress perfusion images are shown for a subject with a thin epicardial fat layer that is present in all three short-axis slices. The proposed approach demonstrated superior performance compared to the established DNN-based analysis approach. The latter resulted in a lower Dice score for basal and mid slices and had a failed segmentation (noncontiguous contour in the septal wall highlighted by the yellow arrow) for the apical slice. Of note, the established approach incorrectly includes the epicardial fat in the segmentation result for the apical slice (highlighted by the magenta arrow). This error is not encountered in the segmentation result corresponding to the proposed approach.

**Supplementary Figure S1**

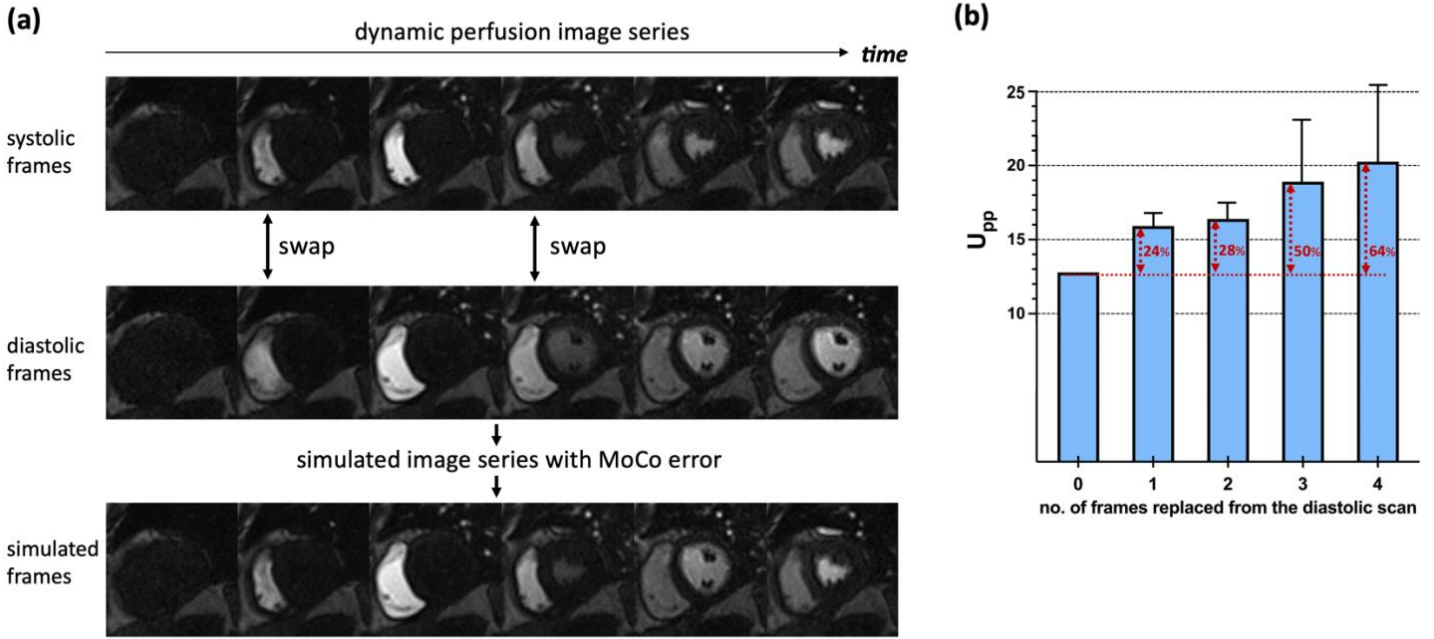

**Supplementary Figure S1 (Additional File 1). Effect of simulated nonrigid motion correction error on the uncertainty quantification metric (Appendix D).** To verify that our per-pixel uncertainty metric behaves as expected in presence of poor motion correction (MoCo) quality, we carried out the following two steps in an example stress perfusion CMR case: **(a)** Generating simulated perfusion image series with imperfect MoCo: as described in Panel (a), this was accomplished by replacing a small subset of systolic frames (mid slice) with diastolic frames (basal slice) from the same scan to create a simulated image series with poor MoCo quality while keeping the temporal dynamics consistent with a typical first-pass perfusion scan; specifically, a total of 30 Monte Carlo simulations were performed where $f$ = 0,1,…,4 time frames ($f$ = 2 in the example shown here) from the diastolic image series (basal slice) replaced the corresponding systolic frames (mid slice). **(b)** The mean per-pixel uncertainty ($U_{pp}$ averaged over 30 Monte Carlo simulations) is plotted; as expected, the uncertainty metric increases as MoCo quality deteriorates.

**Supplementary Figure S2**

(a)

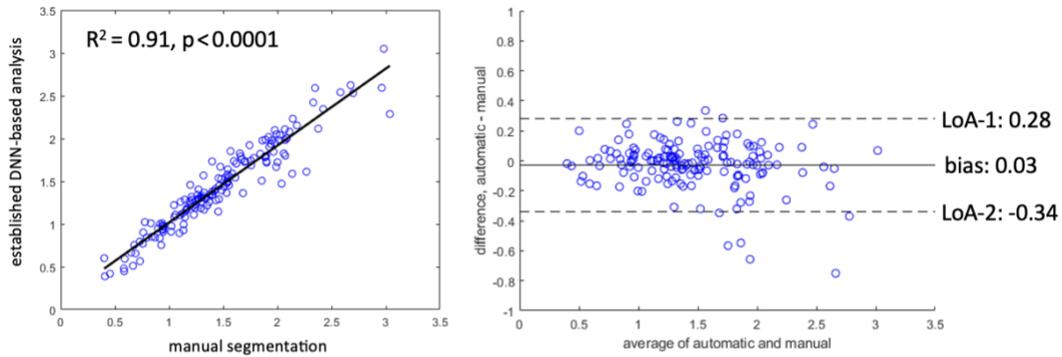

(b)

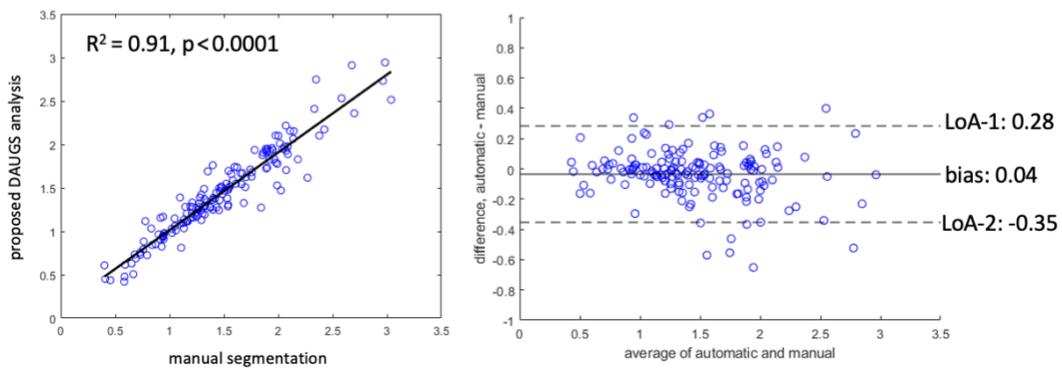

**Supplementary Figure S2 (Additional File 2). Myocardial blood flow quantification results for the internal test set, inD-test (Appendix E).** For further verification of the segmentation results, we quantified myocardial blood flow (in mL/g/min) using the established look-up-table (LUT)-based [Gd] Fermi-constrained deconvolution in 6-segment myocardial divisions for (a) established DNN-based analysis and (b) proposed DAUGS analysis. Both methods demonstrated strong agreement with respect to manual analysis (ground-truth MBF numbers) in terms of correlation coefficient (both approaches: $R^2$=0.91, p<0.0001) and the 95% limits of agreement (LoA) for Bland-Altman analysis (established approach: LoA-1: 0.28, LoA-2: -0.34, bias: 0.03; DAUGS analysis: LoA-1: 0.28, LoA-2: -0.35, bias: 0.04).

**Supplementary Figure S3**

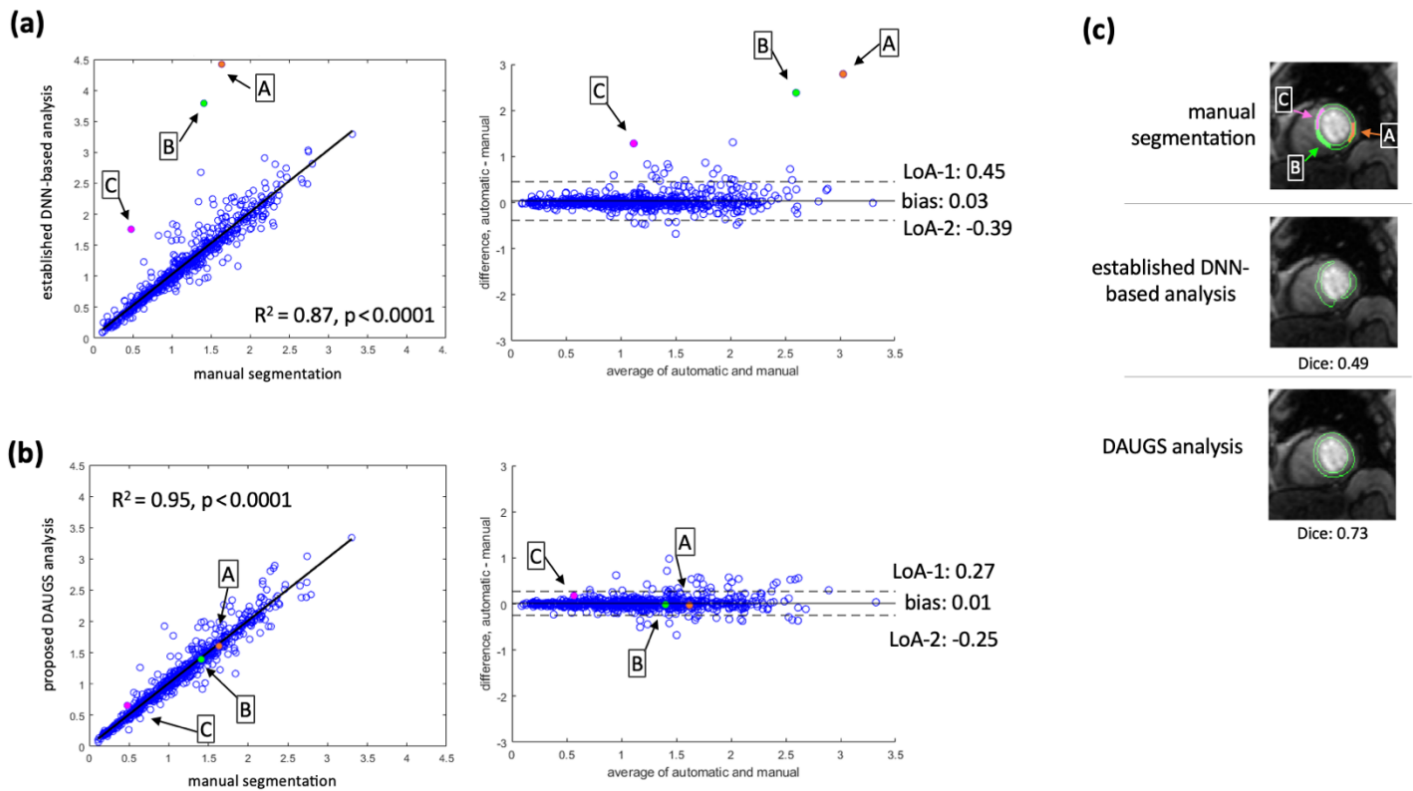

**Supplementary Figure S3 (Additional File 3). Myocardial blood flow quantification results for an external test set, exD-1 (Appendix E).** MBF results for exD-1 is presented for (a) established DNN-based analysis and (b) proposed DAUGS analysis. DAUGS analysis outperformed the established approach by achieving a stronger correlation (established approach: $R^2$=0.87, p<0.0001; DAUGS analysis: $R^2$=0.95, p<0.0001) and tighter limits of agreement in the Bland-Altman plots with respect to manual analysis (established approach: LoA-1: 0.45, LoA-2: -0.39, bias: 0.03; DAUGS analysis: LoA-1: 0.27, LoA-2: -0.25, bias: 0.01). **(c)** Segmentation results for three outliers (3 myocardial segments belonging to the same myocardial slice) that are highlighted as points A, B, and C in (a) and (b) as well.

**Supplementary Figure S4**

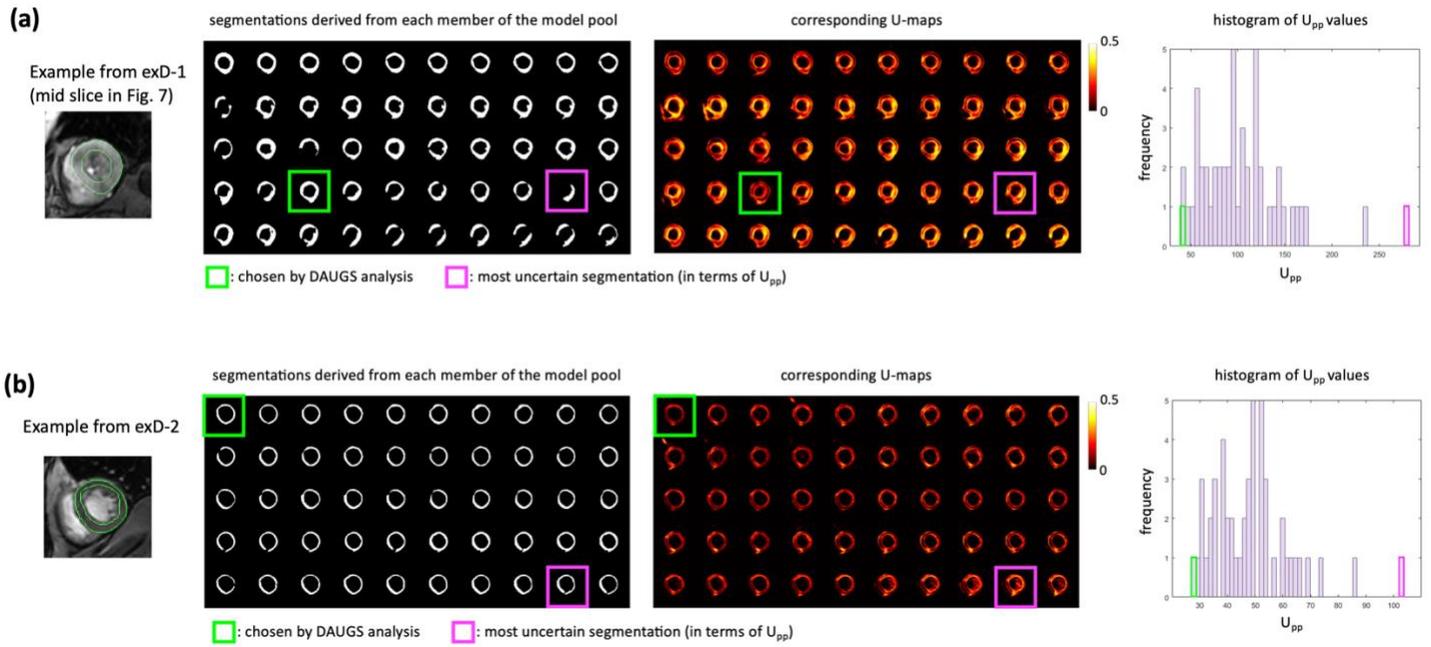

**Supplementary Figure S4 (Additional File 4). Examining the heterogeneity of segmentation solutions in the DNN model pool for two example dynamic image series (Appendix F).** (a) and (b) correspond to the segmentation solutions in the model pool for the proposed DAUGS analysis approach (a total of 50 DNN models as described in Fig. 1) corresponding to two external patient studies: one from exD-1 (same patient as Fig. 7) and one from exD-2. Each panel also includes the corresponding U-maps for each of the 50 segmentation solutions (middle) and show the distribution (histogram) of the uncertainty metric $U_{pp}$ (right side). In the 5x10 matrix of segmentation solutions (and the corresponding U-maps), each row corresponds to one of the 5 training runs/sessions and each column corresponds to one of the 10 checkpoints (i.e., 10 snapshots of the working DNN model during the training process on the internal dataset). These observations support the notion that there is noticeable heterogeneity/diversity (including failed segmentations) among the 50 members of the model pool when tested on the external datasets.

**Supplementary Figure S5**

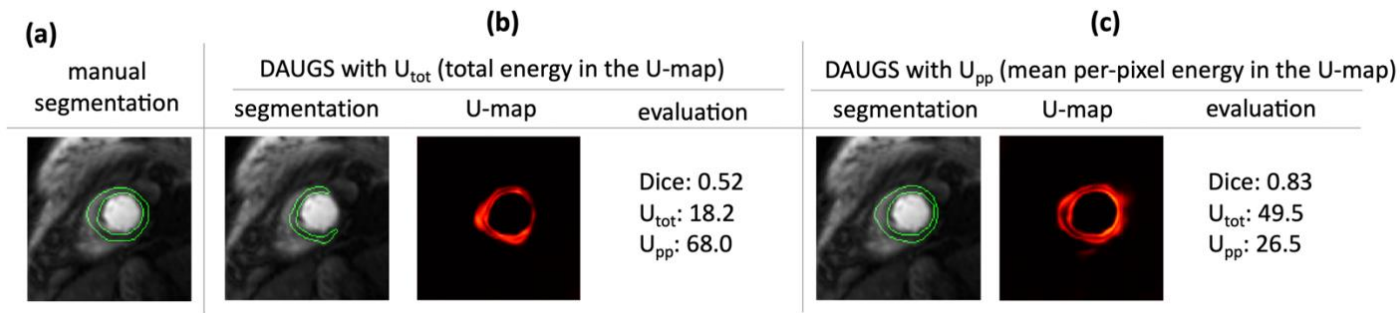

**Supplementary Figure S5 (Additional File 5). Uncertainty-guided model selection with total U-map energy ($U_{tot}$) vs. mean per-pixel energy, $U_{pp}$ (Appendix G).** **(a)** Example apical slice from the external dataset exD-1 with manual segmentation; **(b)** DAUGS analysis results using the alternative uncertainty metric ($U_{tot}$ instead of $U_{pp}$) which results in noncontiguous contours ($U_{tot}$ is minimized for this selected solution but $U_{pp}$ is not); **(c)** DAUGS analysis results using the proposed uncertainty metric ($U_{pp}$) which results in contiguous contours and a higher Dice score ($U_{pp}$ is minimized for this selected solution but $U_{tot}$ is not).